\documentclass[12pt]{article} 
\usepackage{graphicx}
\usepackage{caption}
\usepackage{mathrsfs}
\usepackage{float}
\usepackage{subcaption}
\usepackage[utf8]{inputenc}
\usepackage{longtable}
\usepackage{float}
\usepackage{booktabs}
\usepackage[english]{babel}
\usepackage[dvipsnames]{xcolor}
\usepackage{booktabs}
\usepackage{multirow}
\usepackage{ragged2e}
\usepackage{hyperref}
\definecolor{darkblue}{RGB}{0,0,120}
\hypersetup{
      colorlinks=true,
      linkcolor=darkblue,
      filecolor=darkblue,
      citecolor=darkblue,      
      urlcolor=darkblue,
      }

\usepackage[justification=centering]{caption}
 
\newcommand{\norm}[1]{\left\lVert#1\right\rVert}
\usepackage[toc,page]{appendix}
\usepackage{adjustbox}
\usepackage{multirow}
\usepackage{threeparttable, tablefootnote,booktabs, stackengine}
\usepackage{array}
\usepackage{caption}
\usepackage{placeins}

\usepackage{indentfirst}
\usepackage[margin=1in]{geometry}
\usepackage{subcaption}

\usepackage{amsmath,amsthm,amssymb,scrextend}
\usepackage{cleveref}
\usepackage{bbm}
\usepackage{mathrsfs}
\usepackage{blindtext}

\usepackage{pdflscape}
\usepackage{longtable}
\usepackage{vcell}
\usepackage{booktabs}
\usepackage{natbib}
 \bibpunct[, ]{(}{)}{;}{a}{}{,}%

\DeclareCaptionLabelFormat{andtable}{#1~#2  \&  \tablename~\thetable}

\usepackage{setspace}
\usepackage{lipsum}
\usepackage{footmisc}

 \usepackage{spverbatim}
\usepackage{float}
\usepackage{graphicx}
\usepackage{titlesec}

\newcolumntype{x}[1]{%
>{\centering\hspace{0pt}}p{#1}}%

\titleformat{\section}
{\normalfont\fontsize{13}{13}\bfseries\itshape}
{}{0em}{
\centering
\MakeUppercase}
[
\vspace{0.5ex}%
] 

\titleformat{\subsection}
{\normalfont\fontsize{13}{13}\bfseries\itshape}
{}{-0.5em}{
}
[
] 

\usepackage{comment}
\usepackage{makecell}  

\begin{document}

\begin{titlepage}

\singlespacing

\title{Product Design Using a Generative Adversarial Network: Incorporating Consumer Preferences and External Data}

\author{
Hui Li\thanks{HKU Business School, University of Hong Kong, Hong Kong. Email: \href{mailto: huil1@hku.hk}{ huil1@hku.hk}. } \vspace*{0.3cm}, \
Jian Ni\thanks{Pamplin College of Business, Virginia Tech. Email: \href{mailto:jiann@vt.edu}{jiann@vt.edu}. } \vspace*{0.3cm}, 
Fangzhu Yang\thanks{Economics Division, Bates White. Email: \href{mailto:Fangzhu517@hotmail.com}{Fangzhu517@hotmail.com}.}}


\maketitle

\begin{abstract}
			\singlespacing
The rise of generative artificial intelligence (AI) has facilitated automated product design but often neglects valuable consumer preference data within companies' internal datasets. Additionally, external sources such as social media and user-generated content (UGC) platforms contain substantial untapped information on product design and consumer preferences, yet remain underutilized. We propose a novel framework that transforms the product design paradigm to be data-driven, automated, and consumer-centric. Our method employs a semi-supervised deep generative architecture that systematically integrates multidimensional consumer preferences and heterogeneous external data. The framework is both generative and preference-aware, enabling companies to produce consumer-aligned designs with enhanced cost efficiency. Our framework trains a specialized predictor model to comprehend consumer preferences and utilizes predicted popularity metrics to guide a continuous conditional generative adversarial network (CcGAN). The trained CcGAN can directionally generate consumer-preferred designs, circumventing the expenditure associated with testing suboptimal candidates. Using external data, our framework offers particular advantages for start-ups or other resource-constrained companies confronting the ``cold-start" problem. We demonstrate the framework's efficacy through an empirical application with a self-operated photography chain, where our model successfully generated superior photo template designs. We also conduct web-based experiments to verify our method and confirm its effectiveness across varying design contexts.
			
\end{abstract}
	
\vspace{0mm}\noindent

\end{titlepage}

\clearpage 
\doublespacing
\pagenumbering{arabic} 
	\newpage

\section{Introduction}

Product design is crucial for firms to attract consumers and succeed. Traditionally, this involves hiring professional designers to create small batches of new designs, which is costly and lacks scalability. To evaluate attractiveness of new designs, companies need to gather consumer feedback through focus groups, surveys, and market tests, such as A/B tests or theme clinics. Theme clinics are usually expensive, with vehicle design tests costing over \$100,000 \citep{burnap2023product}. After collecting feedback, firms analyze it using techniques such as conjoint analysis \citep{Green1990Srinivasan}, choice modeling \citep{McFadden1986}, and preference mapping \citep{Urban1993Hauser}. The insights derived from the analyses are then used to refine designs to maximize market share or profits \citep{Allenby1998Rossi}. This iterative, manual process results in significant delays from concept development to market launch.

The rapid advancement of generative AI has the potential to significantly accelerate the product design process. In the design generation phase, generative AI enables industrial designers to quickly explore a wider array of ideas and develop initial concepts much faster than traditional methods (\citeauthor{McKinsey2023}, \citeyear{McKinsey2023}). For instance, generative AI text-to-image tools like DALL-E, Stable Diffusion, and Midjourney can be used to design clothing of a specific style through iterative prompting.\footnote{\url{https://openai.com/index/dall-e-3/},\url{https://stability.ai/news/stable-diffusion-3},\url{https://www.midjourney.com/home}} 

However, a significant gap exists in current AI applications for product design: the design generation stage lacks systematic integration of large-scale consumer preferences \citep{burnap2023product}. AI tools are typically used either before design generation to analyze consumer data and identify needs or after to predict demand for new designs \citep{Dzyabura2011,Toubia2003adaptive}. During design generation, consumer preferences are either missing or incorporated in a limited, ad hoc manner. Although text-to-image tools enable design generation through iterative prompting, these prompts are not systematically generated and may fail to capture diverse consumer tastes. Additionally, the attractiveness of the generated designs to consumers remains uncertain. Predicting consumer demand occurs after design creation, potentially leading to wasted resources on developing and testing unpopular designs.

In contrast to the lack of incorporating consumer preferences into generative AI applications, rich consumer preference information exists in user-generated content (UGC) and remains underutilized by companies in practice. UGC naturally contains valuable consumer insights, as users voluntarily create content. For instance, social media and UGC sites host numerous user-taken photos in various settings, indicating user preferences for those backgrounds. These images can help understand consumer evaluations of new photo backgrounds and serve as inputs for training generative models to create new photo background designs. Despite the wide availability of UGC data, their usage in marketing is mostly descriptive and focuses on extracting preferences or predicting demand. Prescriptive usage of UGC for product design is scant both in marketing academia and in practice. There is no clear guidance on using unstructured UGC data for direct unstructured product designs.

Additionally, finding a scalable solution to integrate consumer preferences and leverage UGC data poses challenges. Developing a generative AI algorithm can be expensive, potentially unaffordable for many companies. Training such algorithms requires large datasets, which may be unavailable to small or startup companies facing a ``cold-start" problem. Furthermore, even after training, generating new designs can be costly due to the high time costs and prices of generative AI tools (see the literature review for a detailed discussion).

Our study addresses existing gaps by focusing on three key questions: (1) How can consumer preference information be integrated into the product design process to guide generative AI in creating appealing designs? (2) How can firms utilize the rich design and preference data in UGC during the design generation stage, particularly for those facing ``cold-start" issues? (3) How can the design generation process be automated systematically, scalably, and cost-effectively?

We propose a novel framework to transform the product design process in the era of generative AI. Our approach utilizes a semi-supervised deep generative framework based on Generative Adversarial Network (GAN) to automatically and cost-effectively create consumer-preferred product designs. Compared to other generative models such as diffusion models (see the literature review for a detailed comparison), our model offers several advantages: 1) it generates realistic images; 2) it is prompt-free, eliminating the need for textual prompts; 3) it incorporates large-scale consumer preferences; 4) it leverages UGC; 5) it enables fast and cost-effective design generation;\footnote{Our trained model can generate a new design in just 0.01 seconds using a consumer-grade laptop, whereas state-of-the-art diffusion models require at least 10 seconds and rely on 2 to 3 enterprise-grade GPUs.} 6) it is tailored for personalized design generation. 

To illustrate the practical application of our framework, we conduct an empirical study with a self-operated photography chain in China. This company operates photo booths in major city shopping malls, allowing customers to select templates, take photos, and print them. As the photo templates are key products, it is crucial for the company to design appealing ones to attract customers. Traditionally, this would involve hiring professional designers and conducting costly consumer tests for each design. As a start-up with limited resources and internal data, the company faces a ``cold-start" problem. Our framework enables the company to systematically integrate revealed preferences from both internal consumer choices and external UGC sources, facilitating rapid and concurrent design generation and evaluation. This approach transforms the process from a sequential design-then-test method to an integrated design-with-preferences strategy, fundamentally changing how firms can develop consumer-oriented products in the era of generative AI.

The self-operated photography chain provides us internal dataset that contains 2195 consumers who had taken at least one photo between October 2017 and August 2018. For each consumer, we observe her full order history, including her selected photo templates and the photos taken with these chosen templates. The mapping from internal consumers to their chosen templates represents internal consumer preference. We further collect external images from UGC websites. These external images are of similar types to those of the internal images, with humans taking photos in front of various backgrounds. They represent both existing external designs and external consumer preferences.

We apply our proposed framework to this empirical context. The framework includes a predictor model that learns consumer preferences and a generative model that uses these predictions to guide design generation. We present variations of the generative model to highlight the framework's benefits, starting with a deep convolutional GAN (DCGAN) that generates new designs from internal and external existing designs in UGC without incorporating consumer preferences. Next, we integrate consumer preferences into the generative process by adapting continuous conditional GAN (CcGAN) from \cite{ding2020ccgan} to our context. Unlike traditional GANs, CcGAN can directionally generate specific types of images by using image labels as additional inputs to train the generator, making it ideal for incorporating consumer preferences. We first train a predictor model with historical consumer choice data to learn preferences, then use these predicted choice probabilities as additional input labels to train the CcGAN. Importantly, we modify the original CcGAN in \cite{ding2020ccgan} to accommodate the ordinal nature of consumer preference labels. The trained CcGAN is then instructed to generate new product designs with higher choice probabilities. We explore two CcGAN versions: an enhanced model using only internal consumer preferences for training the predictor and an advanced model utilizing both internal and external preferences, representing our final framework version. We show that the advanced model can be easily adapted to create personalized designs for individual consumers.

The performance of the generated templates is evaluated through three methods. First, a distance-based metric is employed to assess the similarity of the generated templates to popular existing ones. Popular templates share common features that consumers favor. We use the Caffe BVLC reference model (\citeauthor{jia2014caffe},\citeyear{jia2014caffe}) to obtain embeddings for both existing and generated templates. We calculate the distance between generated templates and the most popular and unpopular existing templates. A smaller (larger) distance to the most popular (unpopular) templates suggests higher potential popularity. Second, consumer evaluations are gathered through an online survey where participants choose between existing and newly generated templates, simulating a real-world photo booth environment to gauge direct consumer preferences. Lastly, we use the trained predictor to calculate the choice probabilities of the generated templates, providing an analytical measure of their expected popularity based on historical data and consumer behavior patterns. Together, these methods offer a comprehensive evaluation of the templates' performance by considering similarity to established popular templates, direct consumer feedback, and predictive analytics.

The evaluation methods consistently indicate that the advanced model, which integrates both internal and external consumer preferences, performs the best. The designs generated by this model are predicted to be more favored by consumers, as confirmed by the distance-based metric, real consumer feedback, and the predictor model. The external dataset plays a crucial role in this process. It provides valuable training inputs for the generative models, enabling them to incorporate new product features from external designs and address the limitation of style variations in the internal data. Additionally, the rich external preferences allow the predictor model to identify popular features from external sources, guiding the generative model to create novel and attractive designs that are absent in the internal data. Two key insights emerge from the analyses: First, relying solely on internal preferences is insufficient. Incorporating external preferences is crucial as they offer additional insights into how consumers evaluate external designs. Without this information, the generative model may fail to integrate popular features from external designs that are missing in internal ones. Second, adapting the CcGAN model to acknowledge the ordinal nature of consumer preference labels is essential; without this adaptation, the generated designs perform worse.

We further conduct a personalized design generation exercise. We randomly select an internal consumer  and use the predictor model to estimate his choice probabilities, termed personalized popularity labels. We use these personalized labels to re-train the general advanced model. This personalized CcGAN is used to generate new templates tailored to the consumer's preferences. The personalized CcGAN outperforms the general CcGANs, demonstrating our model’s capability of conducting personalized design generation.

Our study bridges quantitative marketing research with generative AI advancements. While marketing literature has focused on understanding consumer preferences to guide human designers \citep{Hauser1988Clausing, Griffin1993}, and computer science has developed generative capabilities without systematic preference integration \citep{brock2018large}, our work merges these approaches to create a framework that is both generative and preference-aware. This integration enables product design processes to be data-driven, automated, and consumer-centric, establishing a new paradigm that transforms how firms make product design decisions.

Managerially, our work offers an automated product design solution that is particularly beneficial for smaller start-up firms facing cold-start issues and budget constraints. Firms typically possess internal data regarding their existing product designs and consumer choices. Our framework allows them to systematically, scalably, and cost-effectively integrate product design ideas and consumer preferences from external websites. It also dynamically evolves as more internal and external data are incorporated, enabling firms to quickly adapt to shifts in consumer demand.

The framework transforms traditional product design in three key ways: First, it provides a superior alternative to industry practices relying solely on pure generative models, which ignore consumer tastes during design generation, necessitating ex post facto screening of unpopular designs. Our method generates designs preferred by consumers ex ante, reducing time and resources spent on less appealing options. Second, it offers an effective solution for product design ideation, moving beyond reliance on human designers with limited capacity. By leveraging UGC, it captures real-time preferences from a broad user base. Third, it provides a cost-effective way to obtain consumer evaluations of new designs by using UGC as a source of self-revealed and dynamic consumer preferences, contrasting with traditional costly surveys. This approach is applicable in other settings, such as video meeting backgrounds, where consumer preference information and external resources are abundant.

\subsection{Literature Review}
\label{sec:LR}

Our paper contributes to the literature on the usage of UGC in marketing. Most
of these studies were descriptive and focused on analyzing UGC to extract information and
gain insights for firm operations. Some of them have analyzed textual content.
For example, \citet{netzer2012mine} extracted brand associations from
UGC on forums. \citet{culotta2016mining} extracted the social connections of brands
on Twitter. Other works analyzed visual content (\citeauthor{zhang2023},
\citeyear{zhang2023}; \citeauthor{Zhang2017HowMI}, \citeyear{Zhang2017HowMI};
\citeauthor{pavlov2019increasing}, \citeyear{pavlov2019increasing})
. For instance, \citet{liu2020visual} analyzed consumer-generated
images on social media to study the brand perceptions of consumers. We contribute
to the literature by going one step further, from descriptive to prescriptive work.
Our generative model not only extracts information from unstructured UGC but also uses
this information to make unstructured product design decisions for firms.

We contribute to the literature on deep learning models in marketing.
Most studies in this area use unstructured data to generate structured
firm insights. For instance, \citet{liu2020visual} collected unstructured
image data from social media and used a deep convolutional neural network (CNN)
to extract and predict consumer perceptions of brand attributes. Our
work differs in that we employ unstructured data to directly generate
unstructured firm decisions (i.e., product designs as images)
. Our proposed framework goes beyond the descriptive task
and provides prescriptive solutions for the product design decisions of firms.

We also contribute to the nascent literature on deep generative models
in marketing. A few works have used deep generative
models to create new product designs. For example, \citeauthor{dew2022letting}
\citeyear{dew2022letting} used multiview representation learning to
design brand logos. \citeauthor{burnap2023product} \citeyear{burnap2023product}
proposed a model to augment the commonly used aesthetic design process
by predicting aesthetic scores and automatically generating innovative
and appealing product designs; the predictor model and the generator
models were separately trained. Our work differs in that we incorporate
consumer preference information into the generation process so that
the predictor model guides the generative model. We also introduce and adapt
the CcGAN model to the marketing field, allowing us to systematically
incorporate consumer preferences in an automated and large-scale manner.
In addition, while previous works generated one aspect of product
designs such as logos (\citeauthor{dew2022letting}, \citeyear{dew2022letting})
or the aesthetic appearances of cars (\citeauthor{burnap2023product},
\citeyear{burnap2023product}), we directly design the entire product
(i.e., the photo templates are the products of the firm). This setting allows us to abstract away from other design factors that might influence consumer choices.

\bigskip
\leftline{\it \underline{Relationship to other generative models}}
\medskip


Our work builds on GAN models, which offer distinct advantages over other generative models like diffusion models. While diffusion models excel in producing high-resolution images with fine details, they are more suited for generating surreal scenes and struggle with practical imagery commonly used for information sharing and creation.\footnote{\url{https://openai.com/index/introducing-4o-image-generation/}} Additionally, diffusion models are computationally intensive and slower due to their iterative denoising process. State-of-the-art diffusion models require at least 10 seconds to generate one image and rely on 2 to 3 enterprise-grade GPUs. Advanced models like GPT‑4o image generation, launched in March 2025, can produce high-quality realistic images but require textual prompts and take 30-60 seconds per image, with some taking up to 2 minutes.\footnote{\url{https://help.openai.com/en/articles/8932459-creating-images-in-chatgpt}} In contrast, GANs generate more realistic images, are less computationally demanding, simpler to implement and tune, and enable faster generation, making them more suitable for our goal of large-scale product design generation.\footnote{\url{https://www.geeksforgeeks.org/generative-adversarial-networks-gans-vs-diffusion-models/}} Our trained model can generate a new design in just 0.01 seconds using a consumer-grade laptop, which is 1,000 times faster than diffusion models. In addition, considering the difference in hardware requirements, the generation cost can be 8,000 to 12,000 times lower.\footnote{Our model can generate an image in 0.01 seconds using a consumer-grade laptop. The cost of using an AWS G4 instance (equivalent to a consumer-grade laptop GPU) is \$1.461 to generate 1 million images. In contrast, diffusion models need 3 enterprise-grade GPUs to generate an image in 10 seconds. The cost for 3* A100 GPUs on AWS is \$12,291 to produce 1 million images. Alternatively, generative AI services like DALL-E 2 charge \$18,000 for 1 million images, while Stable Diffusion 3 costs \$35,000 for the same amount of images. The online appendix provides more calculation details.}

Another advantage of the GAN models lies in their conceptual attractiveness. Continuous diffusion models predominantly operate within non-equilibrium thermodynamic frameworks, utilizing score matching methods that progressively transform noise distributions through iterative denoising processes governed by stochastic differential equations. This paradigm inherently embraces non-equilibrium dynamics, where the generative process follows irreversible trajectories through probability density spaces. In contrast, our CcGAN architecture extends the min-max game-theoretic framework, which fundamentally aligns with the equilibrium-seeking conceptualizations prevalent in marketing and applied microeconomics. This allows our model to converge toward Nash equilibria where generator and discriminator functions achieve optimal strategic responses to each other, mirroring the equilibrium-based analytical constructs that underpin consumer preference modeling and market clearing mechanisms in microeconomics. The theoretical congruence between CcGAN's adversarial optimization and economic equilibrium concept is further corroborated by the self-play theorem proving systems (exemplified by DeepSeek's architecture), which demonstrate that adversarial self-competition frameworks with primitive reward-punishment signals can systematically converge toward optimal solutions through iterative strategic adaptation. This mathematical homology between adversarial generative models and strategic equilibrium-seeking economic agents provides our framework with superior interpretability within established paradigms, while maintaining competitive generative performance compared to diffusion-based alternatives. 

Our paper relates to the literature on image generation incorporating human preference, which is a rapidly growing research area in computer vision. State-of-the-art models mainly use diffusion-based models, as seen in works like \cite{xu2023imagereward} and \cite{wallace2024diffusion}. Our model differs in the following ways. First, existing works focus on text-to-image diffusion models, in which human preference is represented by prompt alignment, or human ratings on how well the generated images align with specific text prompts \citep{xu2023imagereward}. By nature, this type of human preference differs from consumer preferences in marketing, which focus on the mapping between images and consumers’ tastes. Second, human preference in diffusion models only captures the average person's preference and overlooks consumer diversity. For instance, in \cite{xu2023imagereward}, preference is vertical, meaning everyone agrees on image quality rankings. In contrast, marketing highlights horizontal preference, where consumers have varying tastes in image styles. Our model addresses this by capturing diverse preferences among heterogeneous consumers. Importantly, our model is well designed for personalized design generation, which we demonstrate using a personalization exercise. Third, diffusion models usually require textual prompts as inputs, limiting their ability to capture large-scale consumer preferences. Our model directly uses unstructured images from consumer data as inputs without textual prompts. Consumer preference is captured by the mapping between consumers and their chosen images in an unstructured and flexible way. Fourth, diffusion models incorporate human preference through an iterative reward feedback process, which is time-consuming. Our model, based on GANs, uses a min-max operation in training, significantly accelerating design generation without iterative denoising. It reduces both training and inference costs for companies.

\section{Conceptual Framework}

\label{sec:model}

\subsection{Overview of the Conceptual Framework}

\label{sec:model_overview}

We propose a semi-supervised deep generative framework for systematically and cost-effectively generating consumer-preferred product designs. Our model offers three key advantages. First, it enables firms to utilize external product designs from sources like social media and image websites, benefiting startups with limited internal designs. Second, it integrates consumer preferences from a firm's internal data to ensure generated designs align with consumer interests. Third, it incorporates external consumer preference information, allowing firms to access and leverage the abundant consumer insights available in user-generated content (UGC). To demonstrate these three advantages, we present three versions of the generative model. The first is a baseline model that uses product designs from both internal and external sources, showcasing the benefit of leveraging external designs from UGC. The second variant enhances the baseline by incorporating consumer preference data from a firm's internal sources, illustrating how internal preferences can guide the generation of more appealing designs. The third variant is the complete model, integrating product design and consumer preference information from both internal and external sources, thereby demonstrating all three advantages of our framework. This subsection outlines the concepts and workflows of the three models, with implementation details provided in the following subsections.

\begin{figure}[htb!]
\vspace{0.4cm}
\centering 
\caption{
Proposed Machine Learning Model for Generating New Designs
}
\vspace{0.1cm}
\includegraphics[width=0.7\textwidth]{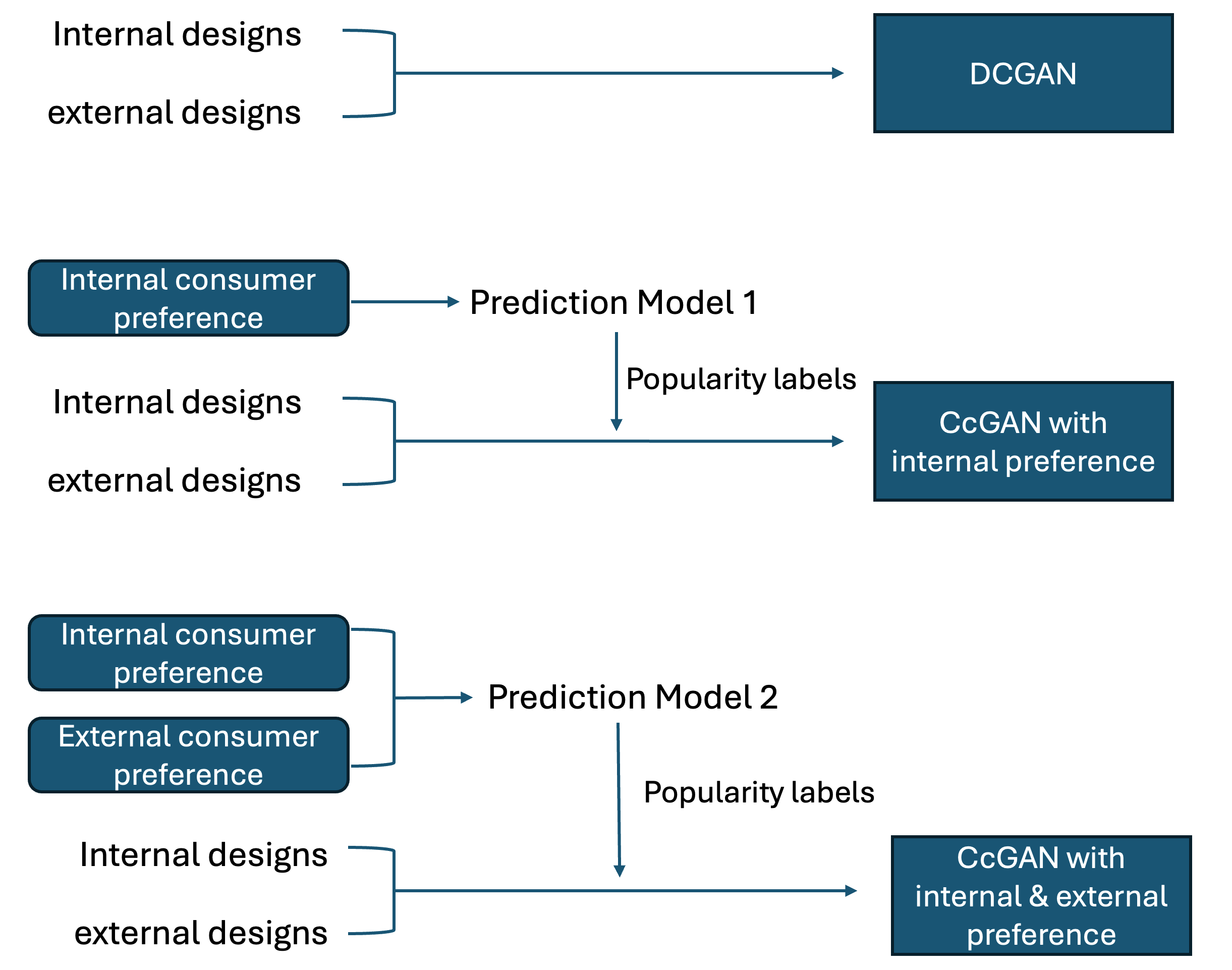}
\label{fig:model}
\end{figure}

\bigskip
\leftline{\it \underline{Baseline model: DCGAN without consumer preferences}}
\medskip

The baseline model uses existing product designs from both internal
and external sources, as shown in Figure \ref{fig:model}. The internal datasets of firms contain existing product
designs that can be used as generative model inputs. However,
these internal designs are usually limited in both quantity and style
variety, especially for start-up firms that face cold-start problems.
To overcome this difficulty, we gather existing product designs from
external sources such as UGC on social media and image websites. UGC
contains a rich set of images and image features. Utilizing UGC as inputs
not only increases the size of the training dataset but also introduces
new image features to the generative process; both of these factors enhance
the performance of the generative model.

Following industry standards, we use GANs (\citeauthor{goodfellow2014generative}, \citeyear{goodfellow2014generative})
to generate new product designs in the baseline model. We choose GAN models over other
generative models such as fully visible belief networks (FVBNs) (\citeauthor{frey1998graphical},
\citeyear{frey1998graphical}) and variational autoencoders (VAEs)
(\citeauthor{kingma2013auto}, \citeyear{kingma2013auto}; \citeauthor{rezende2014stochastic},
\citeyear{rezende2014stochastic}) because GANs were designed
to avoid the many disadvantages associated with other generative models.
As we focus on image synthesis-based tasks, we use the DCGAN (\citeauthor{radford2015unsupervised}, \citeyear{radford2015unsupervised}), which is considered one of the most important steps forward
in terms of designing and training stable GANs.

\bigskip
\leftline{\it \underline{Enhanced model: CcGAN with internal consumer preferences}}
\medskip

Although the baseline model can generate new product designs in an
automated, scalable, and efficient manner, it does not incorporate consumer
preference information. Therefore, the newly generated designs may not
fit consumer needs, and consumers may choose not to buy these new
products. This is a general limitation encountered in most GAN applications in
practice: GANs are used in an engineering sense without incorporating
business insights and consumer needs.

We propose an enhanced generative model that incorporates consumer
preference information into the generative process in an automated and
large-scale manner. To our knowledge, this is one of the first studies to incorporate consumer preferences into a generative
model. Most of the previous studies used consumer preference information
only in prediction tasks. For example, \citet{burnap2023product}
used consumer preference information to train a predictor model and
applied the predictor model to predict how consumers perceived new
product designs; the new product designs, however, were generated without
incorporating consumer preference information. We differ in that we
incorporate consumer preference information jointly in the predictor
model and generator model. In particular, the predictor model
captures consumer preferences and is used to guide the process of designing new
products in the generator model. Therefore, the generator model can
generate new products that are more likely to
be preferred by consumers in a targeted manner.

The internal consumer preference information that we incorporate into the model comes from consumers'
historical product choices in the internal data of firms. Our empirical
application is employed as an example. The internal data of the firm contains the photo
templates that a particular consumer chose and the corresponding photos
taken by the consumer with the chosen templates. Each photo taken
contains both the consumer's appearance and the chosen photo template.
These consumer choices represent the types of product designs that
a particular consumer prefers.

We first train a predictor model to learn the internal preferences of consumers, which we incorporate into the generative process.
The predictor model utilizes the historical photos taken by internal consumers
as inputs. It learns a mapping between a consumer's appearance (i.e.,
her facial traits) and her preferred product designs, assuming that consumers
with similar appearances will prefer similar product designs. To address the high dimensionality of the target problem, we preprocess
the observed consumer faces and product designs using embedding models, obtaining
low-dimensional vector representations for both consumer appearances
and product designs. Using these embeddings as inputs, we train a
deep prediction model to capture high-level abstractions in the prediction
task. We call this predictor model with internal preferences ``Prediction
Model 1,'' as illustrated in Figure \ref{fig:model}. The trained
predictor model can be used to predict the popularity of any given
product design. Specifically, for each pair consisting of a product design and
an internal consumer, the trained predictor calculates the probability
that this consumer will choose the corresponding product. By averaging over the selection
probabilities among all internal consumers, we can obtain a popularity
measure of the particular product.

The trained predictor model is further integrated into the generator
model in a semi-supervised deep learning framework. We employ a conditional GAN
model (\citeauthor{mirza2014conditional}, \citeyear{mirza2014conditional}),
which extends the traditional GAN by enabling the conditional generation
of outputs. Intuitively, the traditional GAN can use a set of dog images
to produce a new set of dog images, but one cannot instruct the
model to create images of a specific type of dog. The conditional GAN
allows one to condition the network on additional information such
as class labels. During training, the conditional GAN takes images
with labels (bulldogs, golden retrievers, etc.) as inputs and learns
the differences between the images with different labels. In this way, it
gains the ability to generate new images with a specific label (a specific
type of dog). 

The conditional GAN is well suited for our task because it allows
us to incorporate consumer preferences and generate new product designs
that are more likely to be chosen by consumers. In our context, we
use the popularity measures derived from the predictor model as additional
inputs (i.e., class labels of product designs). The predicted popularity levels
guide the training procedure of the conditional GAN so that internal consumer
preferences are incorporated into the generative model. Once the generative
model is trained, we can instruct the model to generate new product
designs with higher popularity levels. In this way, the conditional GAN
can generate new designs that are more appealing to consumers.

A key contribution of our proposed framework is that we introduce the conditional GAN to marketing applications and adapt it to suit our context. First, conditional GAN models
typically use categorical labels as additional inputs, while the predicted
popularity measures in our context are continuous class labels. To
address this issue, we employ the CcGAN that was introduced by \citet{ding2020ccgan}. The CcGAN is an
extension of the conditional GAN model, and it was designed to generate images based
on continuous, scalar conditions. By redefining the conventional conditional GAN
approach to make it suitable for continuous scenarios and introducing a
novel method that can integrate continuous labels into the generation process,
the CcGAN outperforms the conditional GAN in terms of effectively handling continuous labels.

Second, rather than using CcGAN from \citet{ding2020ccgan} as-is, we modify it to suit our context by accounting for the ordinal nature of popularity labels. In Ding's applications, the continuous labels represent yaw angles of 3-D chair models and human ages, which do not imply preference (e.g., age 15 is neither better nor worse than age 30). However, in our case, higher popularity levels indicate greater consumer preference; we aim to generate designs with higher popularity. To address this, we adapt CcGAN to recognize the ordinal nature of popularity labels by modifying its loss function to include label weights. Further details are provided in the following subsections. 

\bigskip
\leftline{\it \underline{Advanced model: CcGAN with internal and external consumer preferences}}
\medskip

One limitation of the enhanced model is that the predictor model is
trained only on internal consumer choices and lacks exposure to diverse
design styles that are not present in the internal data. Consequently,
the predictor model may struggle to understand how consumers evaluate
new design styles from the external product designs. Even if the generative
model is trained on both internal and external product designs and
is able to produce ``out-of-the-box'' styles (i.e., styles that
are not present in the internal designs), the predictor model may not
recognize their attractiveness, which limits the performance of the
CcGAN.

Another limitation of the enhanced model is that it does not fully
utilize the information contained in external UGC. By nature, UGC data inherently
contain rich consumer preference information, as users voluntarily
generate the content. In our context of photo templates, the UGC on social
media and image websites usually contains a vast set of photos taken
by individual users in front of various backgrounds. This type of image
represents not only a rich trove of product design data (i.e.,
the backgrounds in the case of photo templates) but also a valuable
source of consumer preference information. The fact that an individual
user chose to take photos in front of a particular background means
that the user likes the background. As such, the external images can also
be used to train the predictor model to understand how consumers evaluate new features in external
product designs.

We propose an advanced model that integrates both internal and external
consumer preferences when generating new product designs. While the previous
studies used external images as extra inputs for training their generative
models (\citeauthor{burnap2023product}, \citeyear{burnap2023product}),
we further exploit the consumer preferences revealed in the external
data to train both the predictive and generative models. Therefore, we can make the most of the available UGC since it contains rich consumer
preference information. Such an approach also saves firms the significant time and money they would have spent on conducting ex post consumer product tests. As discussed in \citet{burnap2023product},
firms often employ A/B tests, or ``theme clinics", to ask consumers
to evaluate different aesthetic designs to improve their product
designs. These theme clinics are costly, as firms usually spend more than
\$100,000 to conduct a single vehicle design test. Therefore, utilizing
consumer preferences derived from external UGC is especially useful for start-up
firms with limited resources. By directly using the existing consumer
preference information contained in UGC, our proposed model significantly
reduces the cost of gathering consumer preference information.

The advanced model follows a workflow similar to that of the enhanced model
except for the fact that the predictor model is trained on both internal and external
consumer preferences. Following the same logic as that used for the internal data,
the predictor model learns the mappings between the users' appearances
and the backgrounds in the UGC images. Upon pooling the mappings between the consumer
appearances and product designs in both the internal and external datasets,
the trained prediction model captures a larger set of consumers and
their preferences over a wider variety of product designs. The predictor
model is labeled ``Prediction Model 2'' in Figure \ref{fig:model}.
Compared with Prediction Model 1, this version can recognize popular out-of-sample 
features from external designs.

The remainder of the workflow is the same as that of the enhanced model.
We use the trained predictor model to compute average choice probabilities
for each product. The average choice probability of a product
serves as its popularity measure and is then used as a continuous
label in the CcGAN, guiding the model to generate popular images. Importantly,
the new predictor model overcomes the limitations of the previous predictor
model, which only recognizes popular product features within internal
consumer preferences. The new popularity measure represents a wider
range of consumer preferences over a wider range of product designs.
Consequently, the CcGAN with both internal and external consumer preferences
is able to generate new product designs that better incorporate the new
features acquired from the external data.

As shown in Figure \ref{fig:model}, the external dataset serves two
purposes. First, the product features obtained from the external data enhance the
training processes of GAN models. This addresses the issue regarding the limited style
variations contained in the internal data by introducing additional style diversity
to train the GAN models to generate improved designs. Second, the
rich trove of external preferences inherent in the external data
guides the GAN models to generate more appealing designs that go beyond
the scope of the internal preferences and designs.

Comparing the three models highlights the advantages of integrating external designs and consumer preferences into product design generation. Our full model, CcGAN with internal and external consumer preferences, is a semi-supervised deep generative framework that automatically and cost-effectively creates consumer-preferred designs. Unlike previous studies that used consumer preference data solely for predictor models, we incorporate this data into both predictor and generator models. Predicted popularity serves as both a label and an input for the generator, enabling it to create products likely to appeal to target consumers. Additionally, we utilize UGC from external sources to enrich the generative model with diverse designs and new features absent from internal data. The extensive consumer preference information in UGC enhances the model's ability to identify popular features not evident in internal consumer data.

\subsection{Prediction Model}

\label{sec:pred_model}

This subsection describes the detailed procedure for training the prediction
model, as shown in Equation \ref{eq:pred_model}. The primary inputs
of the prediction model consist of consumer appearances (i.e., consumer
faces in our empirical context) and product designs (i.e., photo templates
or backgrounds in our empirical context). Given the high dimensionality
of the input data, we preprocess the given images by embedding them into a
lower-dimensional vector space. For consumer faces, we follow
the deep neural network implemented by the OpenFace project (\citeauthor{amos2016openface},
\citeyear{amos2016openface}). As illustrated in the online appendix, this architecture was trained for facial recognition.
It first detects and crops faces from a given image. Then, it produces
a 128-dimensional intermediate layer through a deep neutral network,
which represents a low-dimensional embedding of the face image. For
the product design images, we adopt the Caffe 
BVLC reference model (\citeauthor{jia2014caffe},
\citeyear{jia2014caffe}) to obtain image embeddings. This
model uses the CNN designed by Alex Krizhevsky
(AlexNet) with some modifications, as shown in the online appendix. It provides pretrained general networks that can
be used to obtain low-dimensional embeddings of image characteristics.
Using the pretrained weights of the Caffe BVLC reference model, we
can directly go to fully connected (fc) layer 8 to obtain a 1000-dimensional
intermediate layer that captures the visual characteristics of the
input product design image.

{\footnotesize{}
\begin{equation}
\left.\begin{array}{lll}
\text{Consumer faces} & \xrightarrow{\text{OpenFace}} & \text{128-dimensional vector: }\overrightarrow{X}_{i}\\
\text{Product design images} & \xrightarrow{\text{Caffe BVLC}} & \text{1000-dimensional vector: }\overrightarrow{V}_{j}\\
\text{Other product characteristics} &  & \overrightarrow{Z}_{j}
\end{array}\right\} \text{\quad Concatenated input vector }
\label{eq:pred_model}
\end{equation}
}{\footnotesize\par}

Let the vector $\overrightarrow{X}_{i}$ denote the 128-dimensional embedding
of consumer $i$'s face. Let the vector $\overrightarrow{V}_{j}$ denote
the 1000-dimensional embedding of product design $j$. Let the vector
$\overrightarrow{Z}_{j}$ denote any other low-dimensional product
characteristics that affect consumer's product choices. We concatenate
these three vectors to form the final input vector for training the
predictor model. Let $Y_{ij}$ denote the outcome variable, or the
choice outcome for consumer-product pair $(i,j)$. $Y_{ij}=1$
if consumer $i$ chooses product $j$, and $Y_{ij}=0$ otherwise. The
predictor model is trained to learn the nonlinear relationship between
the input and output variables:
\[
Y_{ij}=f(\overrightarrow{X}_{i},\overrightarrow{V}_{j},\overrightarrow{Z}_{j}|\theta,\varepsilon),
\]
where $\theta$ is the model parameter to be estimated. Since the
prediction task is a classification problem with relatively high-dimensional
inputs, we adopt a random forest to estimate the model parameter $\theta$.
After obtaining the trained predictor, we can use it to calculate
the probability that consumer $i$ will choose product $j$:
\[
p_{ij}(\widehat{\theta})=\mbox{Pr}(Y_{ij}=1|\overrightarrow{X}_{i},\overrightarrow{V}_{j},\overrightarrow{Z}_{j};\widehat{\theta})
\]

\noindent By averaging over the choice probabilities of all consumers
in the sample, we can obtain an aggregate popularity measure for product
$j$: 
\begin{equation}
P_{j}(\widehat{\theta})=\frac{1}{N}\sum_{i\in\mathcal{I}}p_{ij}(\widehat{\theta}),\label{eq:label}
\end{equation}
where $\mathcal{I}$ denotes the total consumer set and $N$ is the
size of the consumer base. The aggregate popularity measure is later
incorporated into the constructed generative models to guide the process of generating new
product designs.\footnote{Using aggregate popularity labels guides generative models to create designs with higher average popularity. Using individual popularity labels, as demonstrated in the personalization section, enables personalized design generation for specific consumers.}

We train two prediction
models: one with internal preferences only and another with
both internal and external preferences. We train these two prediction
models via similar training processes. The difference is that
Prediction Model 1 is trained with internal consumer appearances and
product designs only, while Prediction Model 2 is trained using both
internal and external consumer appearances and product designs. Note
that external consumers are used in the training stage of Prediction Model 2 but not in the prediction stage; when using the trained models to calculate the aggregate popularity
measures, we average over the choice probabilities of the internal consumers
only for both Prediction Models 1 and 2. This is because we are interested
in predicting whether the internal consumers of firms prefer the new template
designs. The external consumer preference information still enhances
the performance of Prediction Model 2 by training the model to recognize
popular design features in external designs. The underlying
assumption is that external consumers with similar appearances to those of the
internal consumers share similar preferences.
Therefore, learning the preferences of the external consumers helps us infer internal consumers' preferences for the new design features from the external designs.

\subsection{Generative Models}

\label{sec:model_gan}

This subsection describes the detailed procedure for training the generative
models.

When training the baseline model without consumer preferences, we adopt
the standard and well-established model architecture of the DCGAN method (\citeauthor{radford2015unsupervised},\citeyear{radford2015unsupervised})
to generate new images. This model takes existing internal and external designs as real image inputs and identifies a set of parameters
that enables the samples generated from random noises to closely resemble
the training data. As the training process is unsupervised and lacks
consumer preference information, the generated images are not necessarily
preferred by consumers.

When training the enhanced model with internal preferences,
we estimate a conditional GAN model and let the internal consumer preferences
guide the generator to produce consumer-preferred designs. Specifically,  we build on the CcGAN model developed by \citet{ding2020ccgan} which is designed for image generation based on continuous labels and utilizes a novel empirical generator loss and a unique label input method. 

Let $M^{r}$ and $M^{g}$ denote the numbers of real and generated
Images, respectively. Let $\boldsymbol{x}\in\mathcal{X}$ denote an image with a size
of $d\times d$. Let $y\in\mathcal{Y}$ denote the corresponding label. We pool together
the existing internal and external product designs to form a group of real
images as input images, yielding a sample size of $M^{r}$. For each
image $j$ in the input group, we use Prediction Model 1 to calculate
its aggregate popularity measure:
\begin{equation}
y_{j}=P_{j}(\widehat{\theta}_{1}),\label{eq:measure}
\end{equation}
where $\widehat{\theta}_{1}$ is the estimated parameter of Prediction
Model 1 and $P_{j}(\widehat{\theta}_{1})$ is defined in Equation
\ref{eq:label}. This popularity measure serves as a continuous class
label and is fed into both the generator and the discriminator in
the CcGAN to guide the generator when creating new designs. Specifically,
building on \citet{ding2020ccgan}, we define the hard vicinal discriminator
loss (HVDL) as
\begin{equation}
\begin{split}\widehat{\mathcal{L}}(D)= & -\frac{C_{1}}{M^{r}}\sum_{j=1}^{M^{r}}\sum_{i=1}^{M^{r}} \mathbb{E}_{\epsilon^{r}\sim N(0,\sigma^{2})}\Big[w_{j}^r \cdot \frac{\mathbbm{1}_{\{|y_{j}^{r}+\epsilon^{r}-y_{i}^{r}|\leq\kappa\}}}{N_{y_{j}^{r}+\epsilon^{r},\kappa}^{r}}\mbox{log}(D(\boldsymbol{x}_{i}^{r},y_{j}^{r}+\epsilon^{r}))\Big]\\
 & -\frac{C_{2}}{M^{g}}\sum_{j=1}^{M^{g}}\sum_{i=1}^{M^{g}}\mathbb{E}_{\epsilon^{g}\sim N(0,\sigma^{2})}\Big[w_{j}^g \cdot \frac{\mathbbm{1}_{\{|y_{j}^{g}+\epsilon^{g}-y_{i}^{g}|\leq\kappa\}}}{N_{y_{j}^{g}+\epsilon^{g},\kappa}^{r}}\mbox{log}(1-D(\boldsymbol{x}_{i}^{g},y_{j}^{g}+\epsilon^{g})) \Big],
\end{split}
\label{eq:lossd}
\end{equation}
where $C_{1}$ and $C_{2}$ are constants,
$\boldsymbol{x}_{i}^{r}$ and $\boldsymbol{x}_{i}^{g}$ stand for
real and generated images, respectively, $y_{i}^{r}$ and $y_{i}^{g}$ are the popularity labels
of $\boldsymbol{x}_{i}^{r}$ and $\boldsymbol{x}_{i}^{g}$, $y_{j}^{r}$ and $y_{j}^{g}$ are target labels in the datasets,
and $\epsilon^{r}$ and $\epsilon^{g}$ are normally distributed random noises. $N_{y_{j}^{r}+\epsilon^{r},k}^{r}$
is the number of the $y_{i}^{r}$ that satisfies $|y_{j}^{r}+\epsilon^{r}-y_{i}^{r}|\leq\kappa$. The label weights \{$w_{j}^r,w_{i}^g$\} are defined to be proportional to the labels as \{$w_{j}^r=ky_{j}^r,w_{j}^g=ky_{j}^g$\}, where $k$ is a hyper-parameter that takes a positive value.
The loss function for the generator is defined as
\begin{equation}
\widehat{\mathcal{L}}(G)=-\frac{1}{M^{g}}\sum_{i=1}^{M^{g}}w_{i}^g \cdot \mathbb{E}_{\epsilon^{g}\sim N(0,\sigma^{2})}\mbox{log}(D(G(\boldsymbol{z}_{i},y_{i}^{g}+\epsilon^{g}),y_{i}^{g}+\epsilon^{g})),\label{eq:lossg}
\end{equation}
where $\boldsymbol{z}_{i}$ is random noise that follows a normal distribution $N(0,\boldsymbol{I})$.
As demonstrated in \citet{ding2020ccgan}, employing the HVDL allows us
to train the discriminator using the images near $y$ rather
then solely relying on images labeled $y$. This approach addresses
the scarcity of real images for each label value. Additionally, we adopt the innovative input method in \citet{ding2020ccgan} to incorporate labels into the model. Specifically, in the generator, we add the labels to the feature map of the first linear layer in an elementwise manner. In the discriminator, we first project the labels to a latent space through an additional linear layer. We then integrate these embedded labels into the discriminator via label projection.

An important modification we made to \citet{ding2020ccgan} is the introduction of label weights, $\{w_{j}^r,w_{j}^g,w_{i}^g$\}, to the loss functions. In Ding's work, the continuous labels, such as yaw angles of 3-D chair models and human ages, do not imply preference. However, in our context, higher popularity levels indicate greater consumer preference, and we aim to generate designs with higher popularity. To enable CcGAN to recognize the ordinal nature of popularity labels, we include label weights, with higher popularity labels assigned larger weights, indicating their greater importance for the algorithm. As demonstrated in the result sections, incorporating these label weights significantly enhances the performance of the generative model.

In summary, combining Equations \ref{eq:measure}, \ref{eq:lossd},
and \ref{eq:lossg}, we integrate consumer preferences into the generation
process by directly incorporating the popularity measure $P_{j}(\widehat{\theta})$
from the predictor model into the loss functions of both the discriminator
and the generator of the CcGAN. This setup allows us to instruct the generator
to create new images with a specific popularity level.

The advanced model incorporating both internal and external preferences is trained using a procedure similar to that of the enhanced model. The only difference is that we use the aggregate popularity measures from
Prediction Model 2, $P_{j}(\widehat{\theta}_{2})$, as additional
input labels to be fed into the generator and the discriminator. As Prediction
Model 2 is trained on both internal and external consumer choices,
the new CcGAN is able to incorporate new design features from the external
sources that are appealing to the internal consumers.

\section{Institutional Setting and Data}\label{sec:institutional}

\subsection{Institutional Background}

\begin{figure}[htbp!]
    \vspace{0.4cm}
    \centering
    \caption{Interface of the Self-Operated Photography Booth}
    \vspace{0.1cm}
    \includegraphics[width=0.8\textwidth]{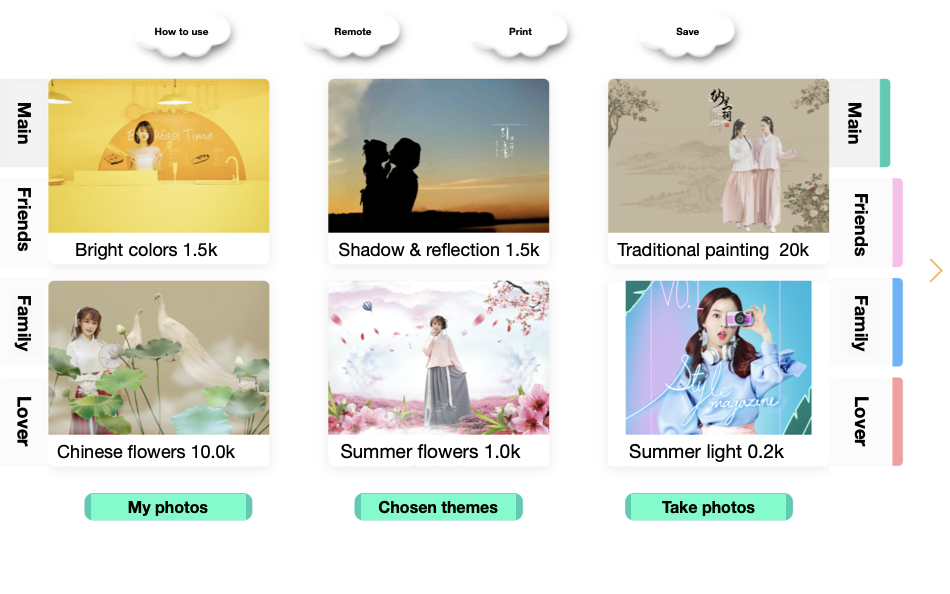}
    \label{fig:interface}
\end{figure}

As an empirical application, we apply our proposed framework to a real business setting: a self-operated photography chain in China. The company operates self-operated photo booths in approximately twenty shopping malls in major cities. The photo booths allow consumers to choose photo templates, take photos with the chosen templates, and print their photos. The photo booths are self-operated, so consumers follow the instructions on the photo booth machines. Figure \ref{fig:interface} shows the interface of such a photo machine. Four photo categories can be selected: main (single), friends, family, and lovers. We focus on the ``main" category in our analysis, which is designed for one consumer to take photos.

\begin{figure}[!htb]
    \vspace{0.4cm}
    \centering
    \caption{Examples of Photo Templates}
    \vspace{0.1cm}
    \begin{tabular}{cc}
    \includegraphics[width=.4\textwidth]{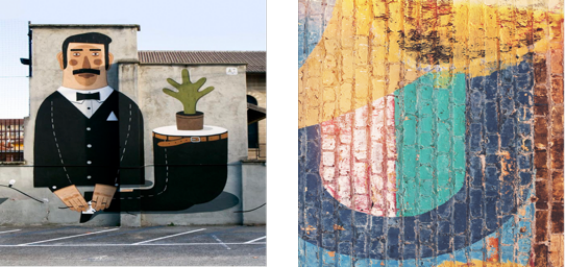} & 
    \includegraphics[width=.4\textwidth]{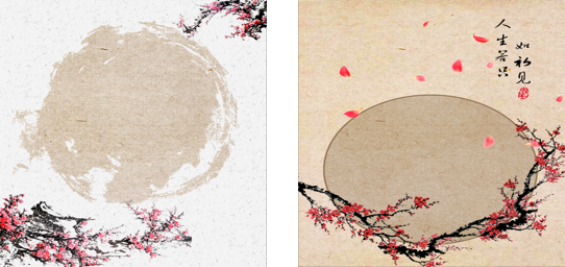} \\
 {\small (a) Graffiti Theme} & {\small (b) Traditional Chinese Theme} \\ 
    \end{tabular}
    \label{fig:template}
\end{figure}

The decision process of the consumers is as follows. Upon entering the photo booth, the consumers are asked to choose photo themes. Several pages of themes are available for the consumers to choose from. The themes range from general topics such as ``modern" and ``classic" to very specific topics such as ``fruits" and ``flowers". The company constantly introduces new themes and ranks these themes by recency, with the more recent themes placed on the first pages on the screen. After clicking on a specific theme, the consumers are presented with all templates under this theme in a pop-out window, and they choose the template(s) with which they will take photos. Each template is typically a combination of a background and a foreground without human subjects involved. For example, Figure \ref{fig:template} shows four photo templates, two from the ``traditional Chinese" theme and two from the ``graffiti" theme. The templates within a specific theme share a similar style, while templates across themes can differ substantially from each other. After choosing the templates, consumers pay for a time window (10, 30, or 60 minutes) and use the time to take photos with the chosen templates. The company charges consumers based on the time they use to take photos, not the number of templates chosen. Consumers can choose an unlimited number of templates and take photos as long as they finish within the paid time window.

As it is in its early development stage, the company does not have a systematic and cost-effective way to design new themes or new templates due to internal resource limitations. The company hires outside designers to create new designs without incorporating consumers' aesthetic preferences. The process of ideating the new themes and designs is not systematic, either, except that holiday-related themes are introduced during holiday seasons. The company manages existing and new designs in a relatively unsophisticated and ad hoc manner: they introduce new designs at a constant pace and remove existing themes that turn out to be unpopular. Our proposed framework can help the company generate new designs in a systematic, cost-effective, and automated way. In particular, the generation process incorporates consumers' heterogeneous preferences and leverages valuable designs and preference information derived from external sources.

\subsection{Internal Data}

The internal dataset obtained from the company contains two parts. The first part includes the original template images that the company offered to customers, without the presence of consumers inside the images. The company offered 83 photo themes in total during the observation period. Each theme contains 1 to 13 templates, yielding a total of 585 templates across all the themes. The second part of the internal dataset contains the complete order histories of 2195 consumers who had taken at least one photo between October 2017 and August 2018. For each consumer, we observe their selected themes and templates and the photos taken with the chosen templates. We also observe the ranking orders of the themes offered on the screen at the times they made their choices.

The internal data patterns suggest that the templates vary significantly in popularity both across themes and within the same theme, potentially due to different design features and heterogeneous consumer preferences. The variation in template popularity, especially the presence of low-popularity templates, suggests that the company needs to improve the attractiveness of its template designs. As another piece of evidence, the internal dataset shows that only 2195 out of the 6038 registered consumers ended up taking photos in the booth after browsing the photo template options. Among the 2195 consumers, fewer than 100 consumers made multiple store visits, and only 15\% of the consumers had repeated time purchases. These numbers suggest that the company can benefit from incorporating consumer preferences to generate better designs that are appealing to consumers.

\subsection{External Data}

To leverage the existing designs and consumer preference information derived from external sources, we further collect an external dataset from two main sources: social media and online image websites. Social media reflects various aspects of consumer tastes, whereas image websites contain a variety of photos with different themes. We collect specific types of photos from these websites, i.e., photos taken by individual users in front of specific types of backgrounds, which are similar to the photo-taking setting in our case. The backgrounds in these external photos can serve as additional design templates for training the generative model. The mappings between the individuals and the backgrounds in these photos can serve as additional consumer preference information to train the prediction model.

\begin{figure}[!htb]
    \vspace{0.4cm}
    \centering
    \caption{Examples of Graffiti Templates: Internal vs. External}
    \vspace{0.1cm}
    \begin{tabular}{cc}
    \includegraphics[width=.45\textwidth]{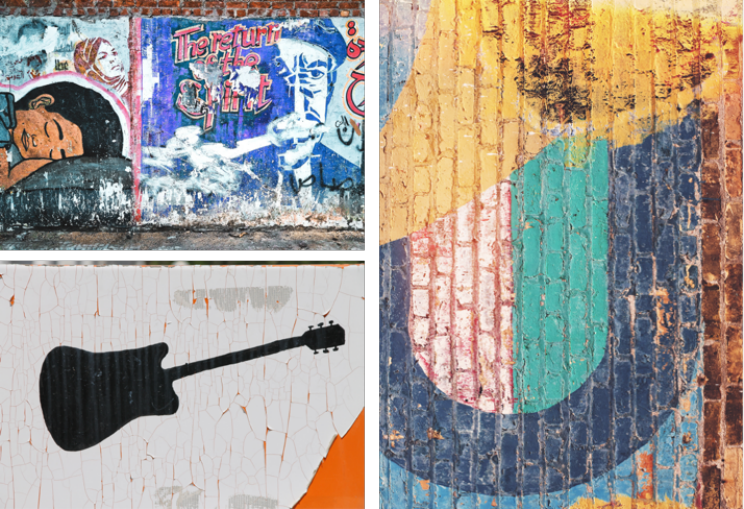} & 
    \includegraphics[width=.4\textwidth]{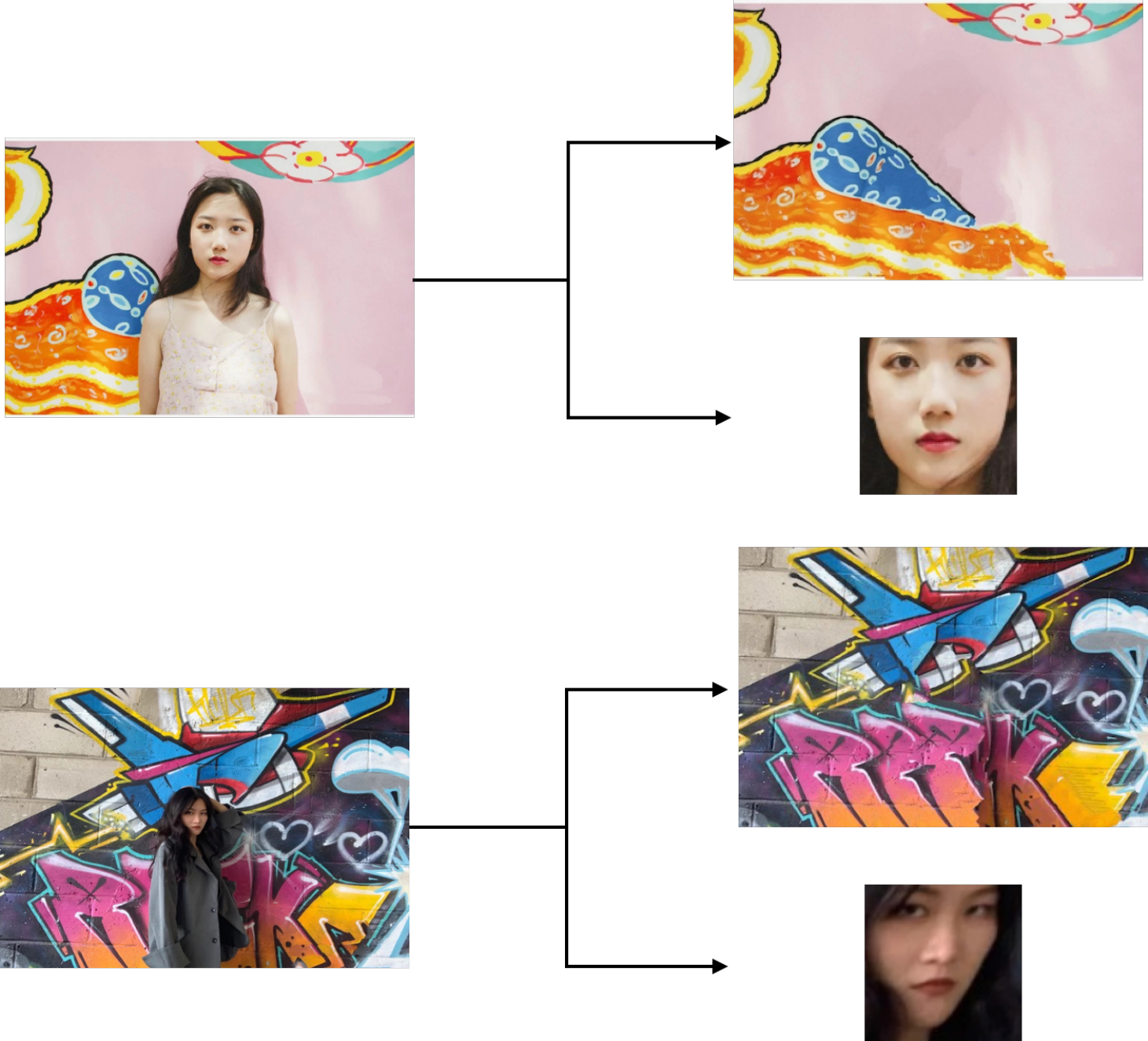} \\
 (a) Internal images & (b) External images (raw, inpainted, cropped) \\ 
    \end{tabular}
    \label{fig:graffiti_example}
\end{figure}

After gathering the external images with human objects and backgrounds, we preprocess the images before they are used as inputs for the proposed model. Specifically, training the model requires a set of pure templates without human objects and a set of consumer face images, while the raw external images we collect are combinations of template backgrounds and consumer faces, so we need to separate the two components. For each external image, we remove the human object in it and inpaint the remaining part of the image to obtain a complete background image, which is similar to the pure template images in the internal dataset. Additionally, we save the cropped human faces as new images, which are similar to the consumer face images in the internal dataset. Figure \ref{fig:graffiti_example} presents examples of the internal templates, raw external photos, inpainted external templates, and cropped consumer faces, all from the graffiti theme. As shown in the graphs, the collected external images are similar in style to the internal images and have sufficient variations. The inpainting process works well and can generate high-quality templates and consumer face images.

\section{Applying Model to Practice}

We apply the proposed model to generate new product designs for the
unmanned photo gallery firm. This section describes the implementation
details.

\subsection{Training the Prediction Model}
\label{sec:pred_model_practice}

The feature embeddings of consumer
faces and background templates serve as the inputs for the prediction
model. Regarding the consumer face features, most consumers took more than one
photo, so multiple face images are available for each consumer.
We first obtain the embedding of each face image. We then average
across the embeddings of the same consumer to obtain a 128-dimensional
facial vector $\overrightarrow{X}_{i}$ for that consumer. With respect to the
template features, in addition to the 1000-dimensional embeddings of
the visual characteristics $\overrightarrow{V}_{j}$, we also include
the display ranking of each template on the screen $\overrightarrow{Z}_{j}$,
as its display ranking affects its probability of being chosen. Concatenating
$\overrightarrow{X}_{i}$, $\overrightarrow{V}_{j}$, and $\overrightarrow{Z}_{j}$
yields a 1129-dimensional feature input for training the prediction model.

{\footnotesize{}
\begin{equation}
\left.\begin{array}{lll}
\text{Consumer faces} & \xrightarrow{\text{OpenFace}} & \text{128-dimensional vector: }\overrightarrow{X}_{i}\\
\text{Photo templates} & \xrightarrow{\text{Caffe BVLC}} & \text{1000-dimensional vector: }\overrightarrow{V}_{j}\\
\text{Template display rankings} & & \text{scalar: }\overrightarrow{Z}_{j}
\end{array}\right\} \text{\quad 1129-dimensional input vector }
\end{equation}
}{\footnotesize\par}

The output variable of the prediction model is constructed from the consumers'
revealed preferences, namely, their template choices. In our setting,
consumers first choose a theme among all themes; only one template
from each theme is displayed as the cover picture for each theme at
this stage. Once consumers click on the chosen theme, they can browse
all templates in the chosen theme and choose some of the templates
to take photos. We assume that consumers choose among templates they have browsed (i.e., cover-page templates of each theme and all templates within the chosen theme). Each choice observation is a consumer-template pair consisting of one consumer and one template she has browsed. If she chose the template, the observation counts towards a positive sample. If she did not choose, the observation counts towards a negative sample. Consumers make binary choices for each browsed templates and can choose multiple templates since consumers can select
an unlimited number of themes and templates within the paid time window.
The chosen templates belonging to the chosen theme naturally serve as the positive
samples. The unchosen templates belonging to the chosen theme serve as the negative
samples because consumers browse them but decide not to take photos with
them. For the unchosen themes, the cover-page templates count toward
the negative samples because consumers browse them upon choosing the
themes and do not choose them. However, the other templates belonging to the unchosen themes do not count
as negative samples and are not included in the analysis because consumers
do not have a chance to see them. Overall, the output variable $Y_{ij}$
for consumer $i$ and template $j$ is constructed as follows:
\begin{equation}
Y_{ij}=\left\{ \begin{array}{ll}
1 & \text{if template }j\text{ is chosen by consumer }i\\
0 & \text{if template }j\text{ is not chosen but belongs to a chosen theme}\\
0 & \text{if template }j\text{ is the cover-page template of an unchosen theme }
\end{array}\right.\label{eq:consider_prob}
\end{equation}

Given the definitions of the input and output variables, we estimate Prediction Model 1 using internal consumer preferences only. We focus on photo samples from the category ``Main (single)," which includes 931 consumers and 33,804 consumer-template pairs. Each pair consists of one consumer and one template that the consumer has browsed. Out of these 33,804 pairs, 3,629 are positive samples, and 30,175 are negative samples. A positive sample indicates that the consumer chose the template, while a negative sample indicates that the consumer browsed but did not choose the template. Negative samples can include templates from both the unchosen templates of the chosen themes and the cover-page templates of the unchosen themes. It's important to note that a template can appear in both the positive and negative samples, depending on the consumer it is paired with. In our data, every internal template has been chosen by at least one consumer, meaning each internal template is part of both the negative and positive samples. The company offered 83 photo themes and 585 photo templates during the observation period. Therefore, there are 585 unique internal templates in both the positive and negative samples for training the prediction model. 

We retain $40\%$ of the observations as the test set
and use the remaining data for training the model. The training and test sets include 20,282
and 13,522 observations, respectively.
As positive samples only account for 10.7\% of the total sample size, an unbalanced sample problem is encountered. To alleviate this problem, we upsample
the minority class to 18,000 to better match the majority class. We concatenate
the upsampled minority class and the majority class to form the final
training set and train the model using a random forest. The number of
trees in the forest is set to 100. The class weights are set to
be balanced.

We estimate Prediction Model 2 using both internal and external consumer
preferences. Specifically, external UGC data contain photos taken by
individual users in front of various backgrounds. Each photo represents
a user's chosen background and thus serves as a positive sample
for the prediction model. Taking the graffiti theme as an example. There are 1375 external photos in the graffiti them, each representing a unique consumer and a unique external template. There are 1375 unique consumers and 1375 unique external templates. These 1375 consumer-template pairs serve as positive samples. Combining these
with the 33,804 internal observations yields a sample size of 35,179 consumer-template pairs.
We use 70\% of the observations, or 24,625 pairs, as the training
set. Among them, 2,544 are positive samples, and 22,081 are negative
samples. To address the unbalanced sample issue, we upsample the minority
class to 20,000 to align with the majority class. We estimate Prediction
Model 2 using a random forest again, maintaining the same hyperparameters
as those employed in Prediction Model 1.

\subsection{Training the Generative Models and Generating New Designs}

Given the trained prediction model, we proceed to training the generator models and generate new designs. While we train the prediction model using samples from all template themes, we train separate generative models for different themes. For illustrative purposes, we present the generative results for the
graffiti theme, as it is one of the most popular themes in the data.
We present the results for other themes (e.g., the traditional Chinese theme) in the online appendix to show the generalizability of our method to different themes.

To train the DCGAN model, we combine the existing templates acquired from the internal
and external sources, resulting in 1388 training templates for the graffiti
theme, 13 of which are from internal data. We exclude images with extreme height or width values. The final training set contains 1321 images. Since the size of our training data is relatively small and the performance of GANs heavily deteriorates in this case, we adopt the differential augmentation (DiffAugment) method proposed by \citet{zhao2020differentiable}.
As the authors pointed out, the performance of a GAN depends heavily
on the sample size of the input training data because the discriminator
will be simply memorizing the exact training set given the small sample
size. The DiffAugment method imposes various types of differentiable
augmentation on both real and generated samples. It is able to generate
high-fidelity images using only 100 images without pretraining, which
sufficiently trains the model using a relatively small training set.

We produce output images with dimensions of $128\times128$.
The latent dimensionality is set to $100$, as is the standard value
for the DCGAN. We use a batch size of $32$ for training the model, with
the learning rate set to $2\times e^{-4}$, and the sizes of the feature
maps in the generator and discriminator are both 64. To obtain stable
results, we train the DCGAN for $5000$ epochs, implementing
$275,000$ iterations in total.

To train the CcGAN with internal consumer preferences, we start by using Prediction Model 1 to calculate aggregate popularity levels for the 1388 existing templates. This model requires display rankings as inputs, which are only available for internal templates because external templates from UGC websites aren't displayed in the firm's system. We address this issue by simulating rankings for the external templates based on the empirical distribution of internal template rankings. To compare generative models, we normalize each popularity level by dividing it by the maximum value within each model, resulting in normalized levels from 0 to 1, which serve as continuous input labels for the generator.

We use the \textit{HVDL} as the
discriminator loss, where the specific loss functions are defined
in Equations \ref{eq:lossd} and \ref{eq:lossg}. Regarding the hyperparameters
of the HVDL, we use the rule-of-thumb formulae to select them, i.e., $\sigma\approx0.028$
and $\kappa\approx0.192$. The label weight parameter $k=10$. The CcGAN is trained for 20,000 iterations
with a constant learning rate of $10^{-4}$ and a batch size of 64.
The dimensions of the latent GAN are set to 256, and the discriminator
updates twice in one iteration. To solve the small sample size problem,
we adopt the DiffAugment method during the training process, similar to
what we did in the DCGAN case. After training the CcGAN, we
instruct the generator to produce new template designs with different
popularity labels, where label=1 indicates the most-preferred class.

To train the CcGAN with internal and external preferences, we employ Prediction Model 2 to calculate popularity levels for the existing
templates and use the normalized popularity labels as inputs to the generator. The hyperparameters of the HVDL, $\sigma\approx0.019$ and $\kappa\approx0.143$, follow the rule-of-thumb formulae. The label weight parameter $k=10$.

\section{Prediction Model Performance}


This section presents performance tests for our prediction models, which are trained and evaluated on templates from all themes.

\subsection{Predicting consumer choices}

\begin{figure}[htb!]
\vspace{0.4cm}
\caption{Examples of the Predicted Numbers of Template Adoptions}
\vspace{0.1cm}
 \centering \includegraphics[width=0.7\textwidth]{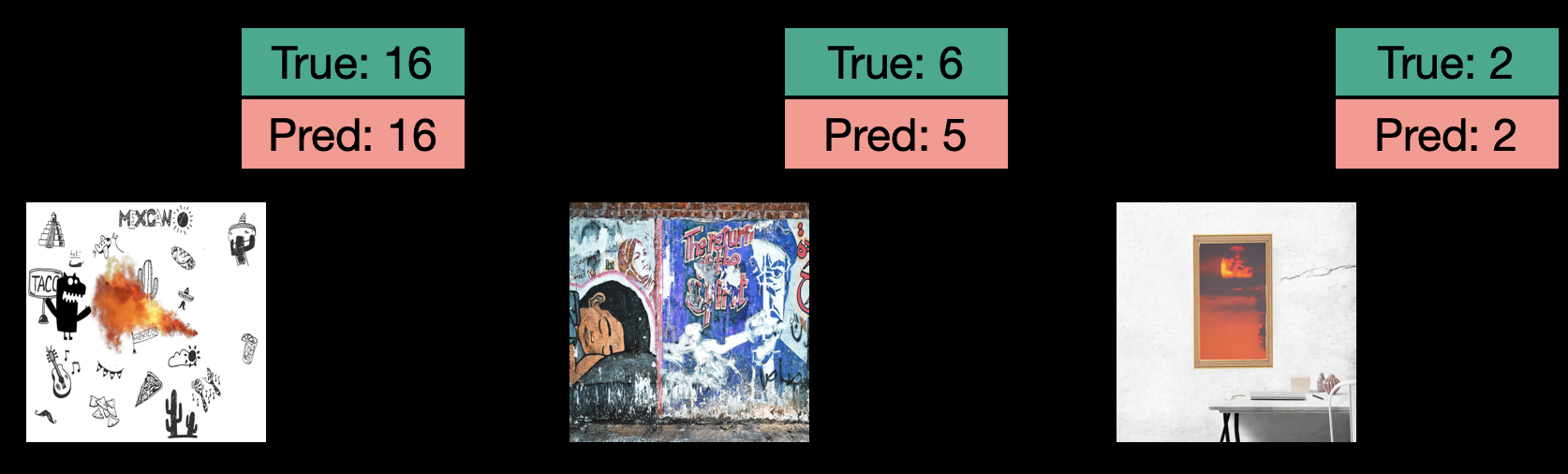}
\label{fig:pred_check} 
\end{figure}

We first show that our prediction model is able to predict observed
consumer choices. Figure \ref{fig:pred_check} presents the observed and predicted numbers
of adoptions for three internal templates in the test set. We can
see that the true and predicted numbers of adoptions are quite close
to each other, suggesting that our prediction model is able to capture
consumer preferences for different template designs. Our model is also able to predict choices of individual consumers, which we show in the online appendix.

\subsection{Predicting design popularity}

\begin{figure}[!htb]
\vspace{0.4cm}
\caption{Comparing the Observed and Predicted Popularity Levels}
\vspace{0.1cm}
 \centering %
\begin{tabular}{>{\centering\arraybackslash}m{2cm}>{\centering\arraybackslash}m{6cm}>{\centering\arraybackslash}m{6cm}}
\hline 
 & Most-Preferred Templates  & Least-Preferred Templates\tabularnewline
\hline 
Observed & \includegraphics[width=0.35\textwidth]{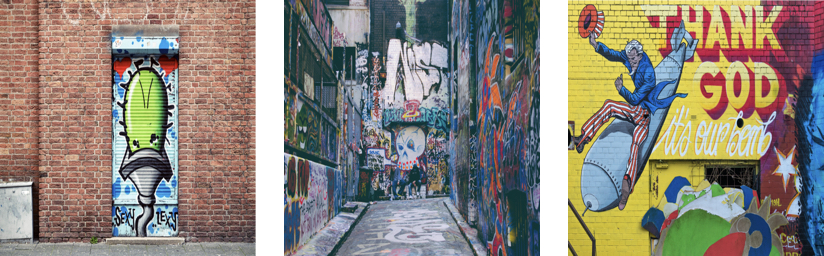}  & \includegraphics[width=0.35\textwidth]{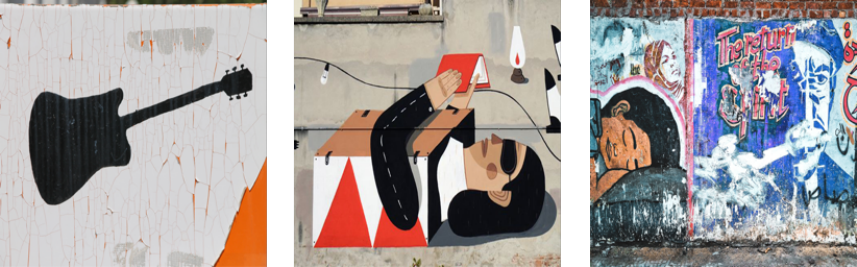}\tabularnewline
\hline 
Prediction Model 1 & \includegraphics[width=0.35\textwidth]{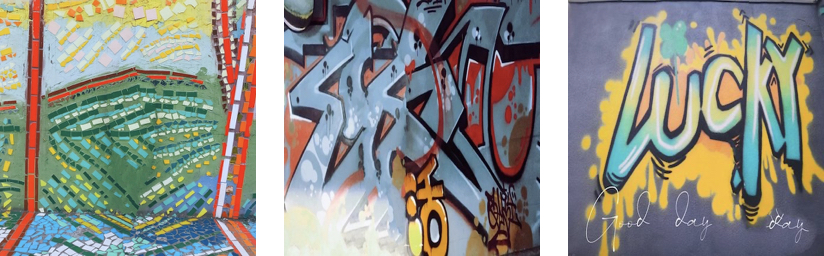}  & \includegraphics[width=0.35\textwidth]{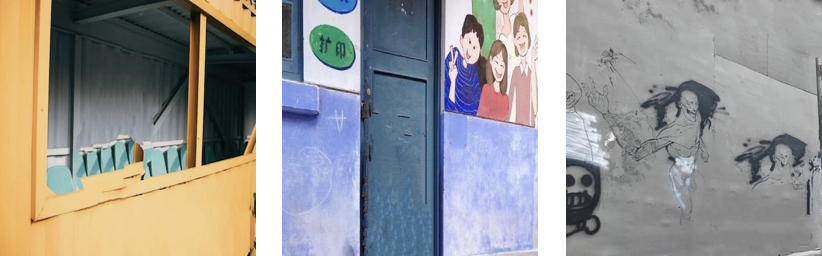}\tabularnewline
\hline 
Prediction Model 2 & \includegraphics[width=0.35\textwidth]{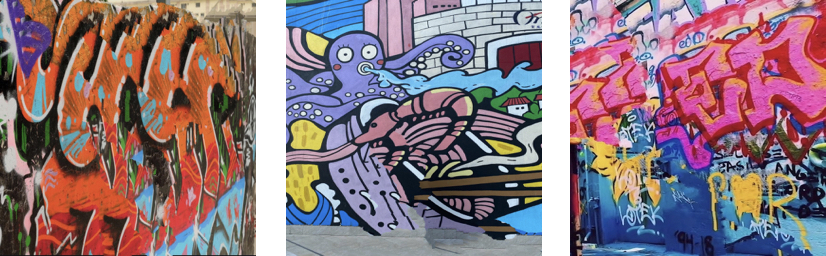}  & \includegraphics[width=0.35\textwidth]{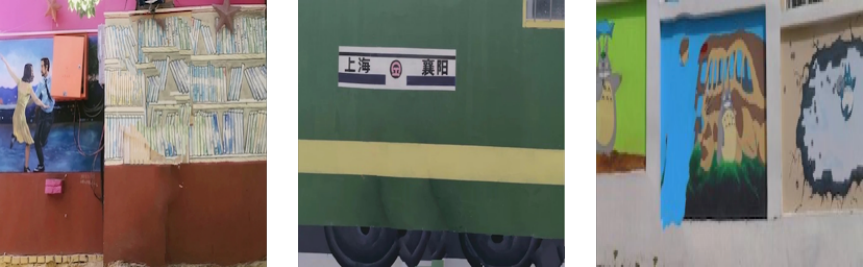}\tabularnewline
\hline 
\end{tabular}
\label{fig:compare} 
\end{figure}

We then conduct a preliminary visual assessment to show that our prediction model is able to predict the popularity levels of different template designs.
We identify the internal templates that are most (least) preferred
by consumers based on their observed numbers of adoptions (row 1 of Figure
\ref{fig:compare}). We then compare their styles with those of the external templates
that our model predicts to have the highest (lowest) average choice
probabilities (row 2 using Prediction Model 1 and row 3 using Prediction
Model 2). We find that the external templates with high (low) predicted
choice probabilities exhibit similar styles to the most (least) popular internal
templates, suggesting that the predicted popularity levels are valid and accurate.

\subsection{Prediction accuracy tests against benchmark models}

Next, we conduct prediction accuracy tests for Prediction Models 1 and 2 on the test set. As shown in Table \ref{tab:performance}, Prediction Model 1 achieves an overall accuracy of $0.793$ and a balanced accuracy of $0.791$. The model's false-negative rate (FNR) the false-positive rate (FPR) are both low, indicating a reasonably high accuracy in predicting consumers' template choices based on facial traits. Incorporating external preferences further improves the model's performance. Prediction Model 2's overall accuracy increases to $0.806$, with a balanced accuracy of $0.825$, and both FNR and FPR are lower. We also compare our method with the logistic lasso model, a common benchmark for handling binary outcomes and high-dimensional (1001+128+1 in our case) input data. Our proposed model outperforms the benchmark, confirming its superiority in predicting consumer template choices.

\begin{table}[h!]
\vspace{0.4cm}
\centering
\caption{Prediction Accuracy Achieved on the Test Set}
\vspace{0.1cm}
\label{tab:performance}
\scalebox{0.85}{
\begin{tabular}{lcccccccccc}
\hline
\multicolumn{1}{c}{\textbf{}} & \multicolumn{4}{c}{\textbf{Prediction Model 1}} & \multicolumn{4}{c}{\textbf{Prediction Model 2}} & \multicolumn{2}{c}{\textbf{Benchmark}} \\
\cmidrule(lr){2-5} \cmidrule(lr){6-9} \cmidrule(lr){10-11} 
\multicolumn{1}{c}{\textbf{Training}} & \multirow{2}{*}{\textbf{Acc.}} & \multicolumn{1}{c}{\textbf{Balanced}} & \multirow{2}{*}{\textbf{FNR}} & \multirow{2}{*}{\textbf{FPR}} & \multirow{2}{*}{\textbf{Acc.}} & \multicolumn{1}{c}{\textbf{Balanced}} & \multirow{2}{*}{\textbf{FNR}} & \multirow{2}{*}{\textbf{FPR}} & \multirow{2}{*}{\textbf{Acc.}} & \multicolumn{1}{c}{\textbf{Balanced}} \\
\multicolumn{1}{c}{\textbf{Input}} & & \textbf{Acc.} & & & & \textbf{Acc.} && & & \textbf{Acc.} \\
\hline
\multicolumn{1}{c}{Internal pref.} & \multirow{2}{*}{0.793} & \multirow{2}{*}{0.791} & \multirow{2}{*}{0.212} & \multirow{2}{*}{0.206} &  &  &  &  & \multirow{2}{*}{0.767} & \multirow{2}{*}{0.789} \\
\multicolumn{1}{c}{only} &&&&&&&&&& \\ 
\multicolumn{1}{c}{Internal and} &  &  &  &  & \multirow{2}{*}{0.806} & \multirow{2}{*}{0.825} & \multirow{2}{*}{0.148} & \multirow{2}{*}{0.202} & \multirow{2}{*}{0.78} & \multirow{2}{*}{0.815} \\
\multicolumn{1}{c}{external pref.} &  &  &  &  &  & & &  & &  \\
\hline
\end{tabular}} \\
\raggedright {\footnotesize{}Note: This table presents the prediction accuracies, balanced accuracy rates, false-negative rates, and false-positive rates for the test set using three models: Prediction Model 1 (trained on internal preferences), Prediction Model 2 (trained on both internal and external preferences), and a benchmark logistic lasso model. The benchmark model is evaluated with internal preferences in the first row and both internal and external preferences in the second row. }{\footnotesize\par}

\end{table}

\section{Generative Model Validity Checks}

\label{sec:result2}

\subsection{Face validity: Generating meaningful designs}

\begin{figure}[!htb]
\vspace{0.4cm}
\caption{Real vs. DCGAN-Generated Images}
\vspace{0.1cm}
 \centering \begin{subfigure}{.30\linewidth} \centering \includegraphics[width=1\linewidth]{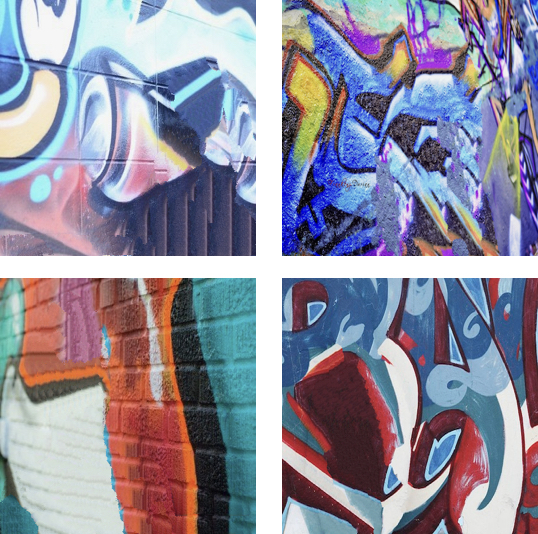}
\caption{{\footnotesize{}Real }}
\end{subfigure} \hspace{3em} \begin{subfigure}{.30\linewidth}
\centering \includegraphics[width=1\linewidth]{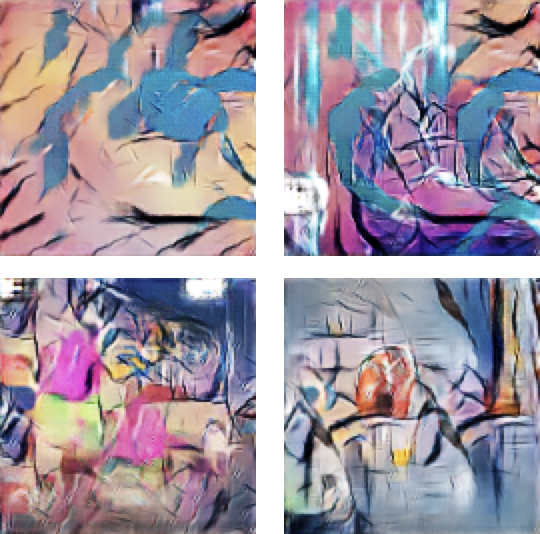}
\caption{{\footnotesize{}DCGAN-Generated }}
\end{subfigure}
\label{fig:gan_graffiti} 
\end{figure}

We first demonstrate the face validity of the generative models by showing
their ability to generate meaningful product designs. Specifically, we need
to show that our new model-generated designs are derived from the same theme
as that of the input designs (i.e., graffiti). We start with a simple visual
assessment. Figure \ref{fig:gan_graffiti} compares the real images
and the new images generated by the trained DCGAN. We can see that
the generated images are similar in style to the real images.
At the same time, 
the generated images exhibit sufficient style variations, which guarantees that no modal collapse occurs during the training process.

In addition to this simple visual examination, we use a pretrained image classifier, ResNet18 (\citeauthor{he2016deep}, \citeyear{he2016deep}), to conduct transfer learning to validate the meaningfulness of our DCGAN-generated images. Intuitively, if the generated images are indeed from the same theme as that of the real images (i.e., graffiti), ResNet18 should be able to classify them into the same category as that of the real images. First, we retrain the classifier to recognize graffiti and nongraffiti images using 100 real graffiti images (both internal and external, with $\text{label}=1$) and 70 real nongraffiti images (internal, with $\text{label}=0$) as inputs. Second, we test the validity of the retrained classifier on a test set consisting of real images. The classifier has an accuracy rate of 97.2\% when classifying the images into graffiti and nongraffiti categories. The false-negative rate and false-positive rate are both very low (2.4\% and 3.4\%) on the test set, further validating the ability of the classifier to predict the graffiti category. Finally, we use this retrained classifier to predict the labels for the set of generated graffiti images and present the hit rate, defined as the percentage of the generated images that are predicted to be in the graffiti category (i.e., $\text{label}=1$). We find that the hit rate is 99\%, meaning that almost all the generated images are classified into the graffiti category by the classifier. This means that our generative model is able to produce meaningful designs with high fidelity.

\subsection{Controllably generating designs with certain popularity levels}

\begin{figure}[h!]
\vspace{0.4cm}
\caption{CcGAN-Generated Images \label{fig:CcGAN-Generated-Images}}
\vspace{0.1cm}
\noindent \begin{centering}
\begin{tabular}{cc}
\multicolumn{2}{c}{CcGAN with Internal Preferences Only}\tabularnewline
\includegraphics[width=0.25\textwidth]{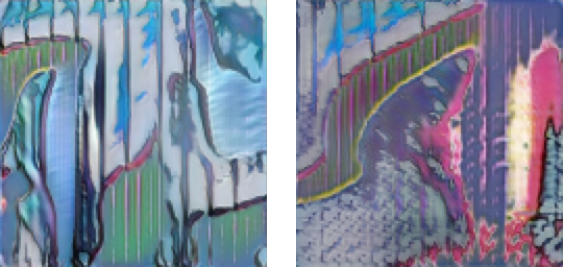} & \includegraphics[width=0.25\textwidth]{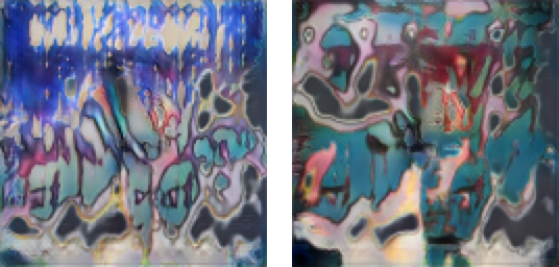}\tabularnewline
(a) Least-Preferred Class & (b) Most-Preferred Class\tabularnewline
 & \tabularnewline
\multicolumn{2}{c}{CcGAN with Internal and External Preferences}\tabularnewline
\includegraphics[width=0.25\textwidth]{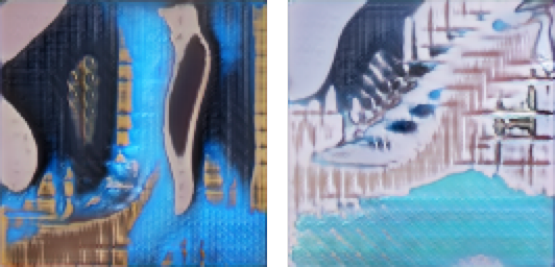} & \includegraphics[width=0.25\textwidth]{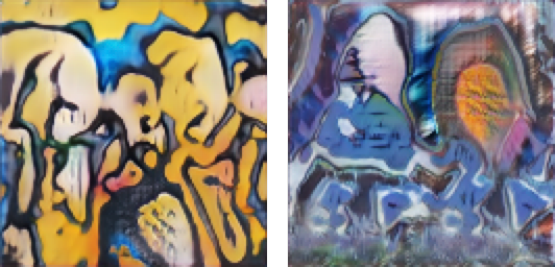}\tabularnewline
(a) Least-Preferred Class & (b) Most-Preferred Class\tabularnewline
 & \tabularnewline
\end{tabular}
\par\end{centering}
\end{figure}

Our proposed framework allows for generating designs with specific popularity levels, unlike traditional GAN models. We use the CcGAN model with internal consumer preferences to generating new designs from the least-preferred and most-preferred classes. The first row of Figure \ref{fig:CcGAN-Generated-Images} shows that both classes produce graffiti-style images, passing the face validity check. The least-preferred class features simpler shapes and patterns, while the most-preferred class has more intricate designs, consistent with internal images.

Additionally, we generate new designs using the CcGAN model with both internal and external preferences. The second row of Figure \ref{fig:CcGAN-Generated-Images} shows more pronounced style differences between the classes. The least-preferred class exhibits simpler colors and shapes, while the most-preferred class displays more intricate and colorful designs, better exemplifying the graffiti theme. This indicates that incorporating external preferences enhances the quality of generated images in the most-preferred class.

\begin{figure}[htb!]
\vspace{0.4cm}
\caption{Generated Graffiti Images with Continuous Labels}
\vspace{0.1cm}
\centering \includegraphics[width=0.99\textwidth]{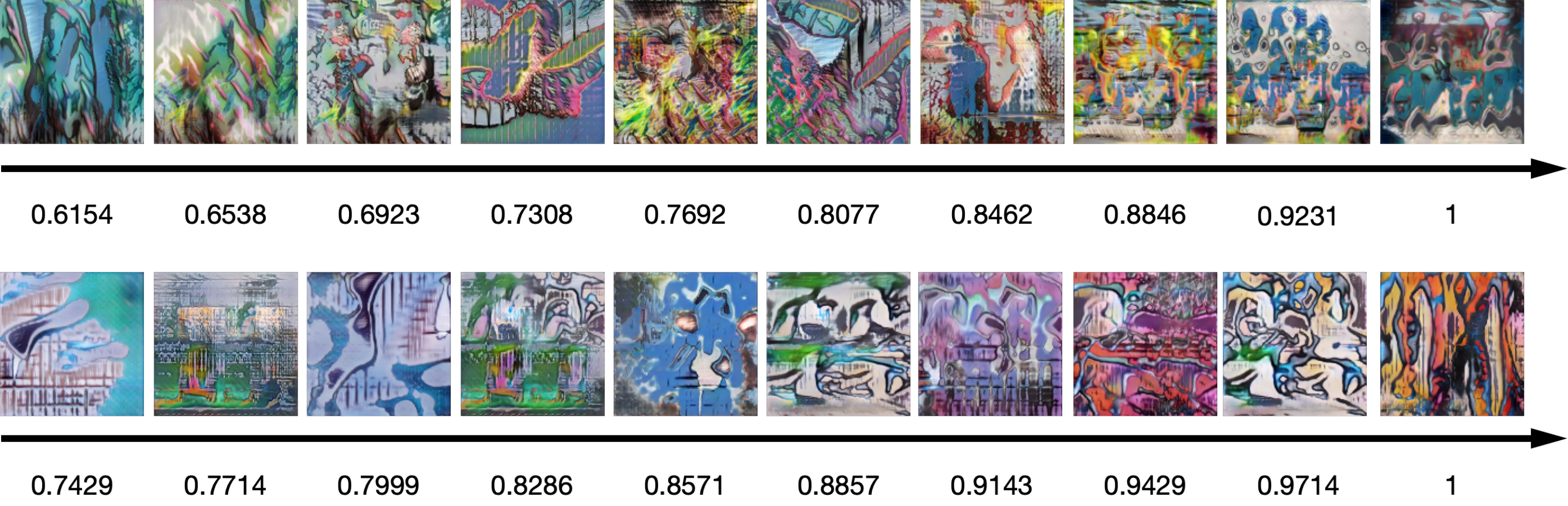}
\label{fig:continuous_label} 

\raggedright {\small{}Note: The top row shows the images generated by the CcGAN with internal
preferences only, while the bottom row shows the images generated by the CcGAN with both internal and external preferences.}{\small\par}
\end{figure}

As our CcGAN models utilize continuous popularity labels, they are
able to generate a range of images with continuous popularity levels.
Figure \ref{fig:continuous_label} displays the progression of the generated
images ranging from the lowest to the highest popularity level. The
top row shows the images generated by CcGAN with the internal preferences,
while the bottom row shows the images generated by CcGAN with both internal
and external preferences. The style of the generated images gradually changes as the popularity level increases. Images with higher popularity levels have more sophisticated
and clearer patterns, especially for the images generated by CcGAN with both preferences.

\section{Generative Model Performance Tests}

\label{sec:result3}

After validating the generative models' capability to controllably generate designs, we now assess their effectiveness in generating designs favored by consumers, which is the primary objective of our framework. We conduct three types of performance tests: a distance-based metric comparing designs to known popular ones, an online consumer survey, and choice probability assessments using prediction models.

\subsection{Distance-based metric}
\label{sec: distance_based_metric}

We propose a distance-based metric as an indicator of popularity. The core idea is that popular templates share common features, suggesting that consumers tend to favor certain types of templates. By measuring the similarity between newly generated images and existing popular templates, we can determine their potential appeal. If the new templates closely resemble the popular ones, they are more likely to be attractive to consumers. Additionally, if they differ significantly from unpopular templates, their chances of being well-received also increase. The metric provides an independent assessment of model performance because it does not rely on any model within our framework. Instead, it leverages the intrinsic similarity among the image features to
suggest popularity.

We start by identifying the most and least popular templates among existing internal and external templates, which serve as the targets for measuring the distances. For internal templates, popularity is determined by the number of consumers selecting each template—the most popular template is chosen by the largest number, while the least popular is chosen by the smallest. For external templates, direct consumer choice data isn't available since each external image corresponds to a unique consumer's choice of a unique template. Instead, we assign a popularity score based on visual similarity to frequently observed patterns in the external dataset. Patterns that are repeatedly seen among external templates indicate popularity among external consumers. To identify these popular patterns, we use the Caffe BVLC reference model to obtain feature embeddings for all external templates to capture their semantic and visual characteristics. These embeddings are clustered using KMeans into ten clusters, with the cluster centers representing common or popular visual archetypes. An external template's popularity score is calculated by computing the cosine similarity between its embedding and each cluster center, averaging these similarities. This score reflects the image's alignment with popular patterns — images with higher scores are more similar to common archetypes and are thus deemed more popular, while those with lower scores deviate from typical patterns.

The calculation procedure of the distance-based metric is as follows. Denote the most popular internal and external templates $j\in\mathcal{J}$ and the least popular internal and external templates $k\in\mathcal{K}$. First, we use the Caffe BVLC reference model to obtain feature embeddings of the new templates generated by different generative models and the most and least popular existing templates. Second, we calculate the distance between each generated template $i$ and each of the most and least popular existing templates. We further normalize the distances to be within the range of {[}0, 1{]}. Let $d_{ij}$ and $d_{ik}$ denote the normalized distances between template $i$ and the most popular template $j$ and the least popular template $k$, respectively. Finally, we calculate the distance-based metric for each generated template and average the results over all templates generated from the same generative model to obtain a measure for each generative model. To summarize, the distance-based metric for each generative
model is defined as
\[
D=\frac{1}{N}\sum_{i=1}^{N}\Big\{\sum_{j\in\mathcal{J}}d_{ij}^{2}+\sum_{k\in\mathcal{K}}(1-d_{ik})^{2}\Big\},
\]
where $N$ denotes the total number of templates generated by each
generative model, $\mathcal{J}$ denotes the set of the most popular internal and external templates, and $\mathcal{K}$ denotes the set of the least popular internal and external templates. This distance-based metric decreases when the generated
templates are more (less) similar to the most (least) popular existing templates. The smaller the distance-based metric is, the better.

We calculate the distance-based metric for four sets of images: 1) the internal original templates; 2) 100 new templates generated by DCGAN; 3) 100 new templates generated by CcGAN with internal preference from the most-preferred class; 4) 100 new templates generated by CcGAN with internal and external preferences from the most-preferred class. Table \ref{table:pred_distance_new} presents the mean and median distance-based
metrics for each set of images.

\begin{table}[htbp!]
\vspace*{0.2cm}

\caption{Distance-Based Metric: Mean and Median}
\label{table:pred_distance_new}
\centering
\scalebox{0.8}{
\begin{tabular}{lccccc}
\hline 
Statistics & \makecell{Internal \\ original images} & \makecell{DCGAN}  & \makecell{CcGAN with \\internal preference} & \makecell{CcGAN with internal  \\ and external preferences \\ (no label weights)} & \makecell{CcGAN with internal \\ and external preferences}  \tabularnewline
\hline 
Mean   & 1.32  & 1.26  & 1.25 & 1.18 & 1.14 \tabularnewline
Median & 1.38  & 1.25  & 1.22 & 1.19 & 1.13 \tabularnewline
\hline 
\end{tabular}}

\raggedright {\footnotesize{}Note: This table presents the mean and median distance-based metrics for images generated by different GAN models. Smaller values indicate higher popularity.}{\footnotesize\par}
\end{table}

We find that CcGAN with internal and external preferences performs the best, followed by CcGAN with internal preference and DCGAN. All generative models outperform the original internal templates provided by the company. Specifically, DCGAN (Column 2 of Table \ref{table:pred_distance_new}) performs the worst among the generative models because it does not account for consumer preferences. CcGAN with internal preference (Column 3 of Table \ref{table:pred_distance_new}) performs better as it is guided by Prediction Model 1, which captures internal preferences, enabling it to generate templates more appealing to internal consumers. However, this improvement is limited because Prediction Model 1 is trained solely on internal preferences and fails to fully recognize the appeal of new features from external templates. Consequently, any novel features from external templates that differ from popular internal features are perceived as unpopular by Prediction Model 1.

CcGAN with both internal and external preferences (Column 5 of Table \ref{table:pred_distance_new}) performs the best by integrating external preference information through Prediction Model 2. This integration allows CcGAN to fully exploit appealing features from external templates. Specifically, Prediction Model 2 learns the types of features consumers prefer in external templates. The underlying assumption is that external consumers who resemble internal consumers have similar preferences, so external consumer preferences can inform those of internal consumers. Consequently, Prediction Model 2 can identify which external features might appeal to internal consumers, guiding CcGAN to integrate popular features from both internal and external sources, leading to the greatest improvement in model performance.

\begin{figure}[h!]
\vspace{0.4cm}
\caption{CcGAN-Generated Images without Label Weights \label{fig:CcGAN-Generated-Images_old}}
\vspace{0.1cm}
\noindent \begin{centering}
\begin{tabular}{cc}
\multicolumn{2}{c}{CcGAN (internal preferences) without label weights}\tabularnewline
\includegraphics[width=0.25\textwidth]{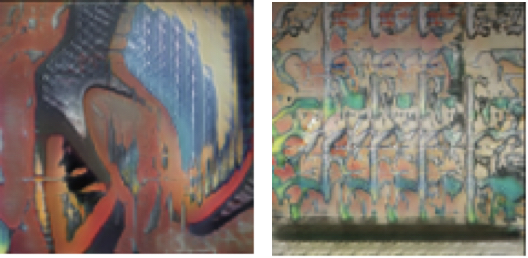} & \includegraphics[width=0.25\textwidth]{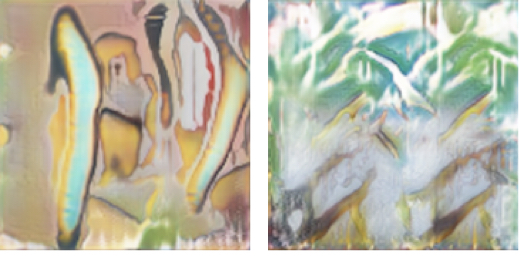}\tabularnewline
(a) Least-Preferred Class & (b) Most-Preferred Class\tabularnewline
 & \tabularnewline
\multicolumn{2}{c}{CcGAN (internal and external preferences) without label weights}\tabularnewline
\includegraphics[width=0.25\textwidth]{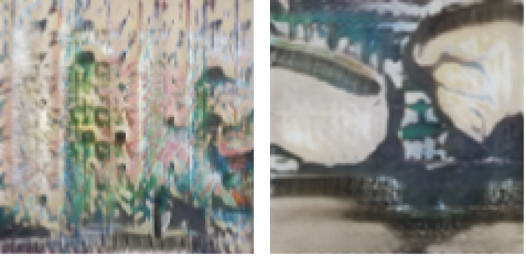} & \includegraphics[width=0.25\textwidth]{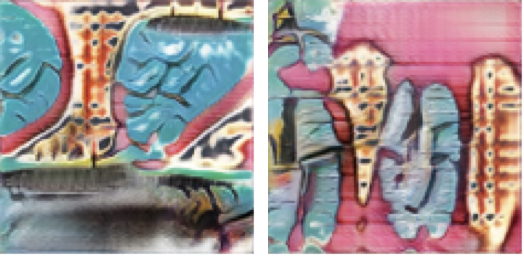}\tabularnewline
(a) Least-Preferred Class & (b) Most-Preferred Class\tabularnewline
 & \tabularnewline
\end{tabular}
\par\end{centering}
\end{figure}

It is noteworthy that the label weights in the loss functions (Equations \ref{eq:lossd} and \ref{eq:lossg}) play a crucial role in enhancing the attractiveness of generated templates. We modify the CcGAN from \cite{ding2020ccgan} by incorporating label weights to account for the ordinal nature of labels in our context. To demonstrate their impact, we trained the CcGAN without label weights. Figure \ref{fig:CcGAN-Generated-Images_old} shows sample templates from this version, which appear less sophisticated and more blurred compared to those in Figure \ref{fig:CcGAN-Generated-Images}. The distance-based metric for this CcGAN without label weights (Column 4 of Table \ref{table:pred_distance_new}) indicates worse performance compared to the full model with label weights (Column 5 of Table \ref{table:pred_distance_new}). These findings underscore that incorporating label weights significantly enhances the quality of the generated images.

Overall, the results reveal three key insights: (1) Integrating consumer preference information enhances performance, with CcGAN consistently outperforming DCGAN. (2) Including external consumer preferences is essential for improving image quality, as it helps the predictor model identify and incorporate appealing external features. (3) Adapting CcGAN to recognize the ordinal nature of labels is crucial for generating high-quality designs.

\subsection{Online consumer survey}

We further obtain direct consumer evaluation on the generated designs through an online survey on Credamo, a leading survey platform in China.\footnote{\url{https://www.credamo.com/\#/dataMarket}} Mimicking a self-operated photo booth, respondents choose from photo templates, including original internal templates and those generated by the three generative models. As there are many original and newly generated templates, we randomly select 4 generated templates per generative model (denote as A1-A4 for CcGAN with internal and external preferences, B1-B4 for CcGAN with internal preference, C1-C4 for DCGAN) and 2 templates from the 13 internal ones (denoted as D1 and D2). This fixed set of templates is used for the survey.

In the survey, respondents are introduced to a self-operated photo booth concept where they select photo templates as backgrounds for pictures. They are informed that their choices will help design popular backgrounds, aiding photo studios in optimizing services. Respondents imagine entering a self-service photo booth to take personal photos. They are presented with 8 templates (similar number of templates per theme are presented in the real business setting) and asked to choose the 2 most suitable backgrounds. The 8 templates include: 2 from CcGAN with internal and external preferences (randomly from A1-A4), 2 from CcGAN with internal preference (randomly from B1-B4), 2 from DCGAN (randomly selected from C1-C4), and 2 internal original templates (D1-D2). The display order is randomized, and the set of templates varies per respondent. Each respondent participates in one choice occasion. The survey includes attention checks and basic demographic questions, and responses failing attention checks are excluded. The final sample includes 288 respondents. A sample survey is provided in the online appendix.

Table \ref{table:online_survey} presents the template-level and model-level choice probabilities from the survey data. Template-level choice probability is calculated as the number of times a template is chosen divided by the number of times it is shown. Model-level choice probability is calculated as the number of times templates from a specific model are chosen divided by the number of times they are shown. We find that CcGAN with internal and external preferences performs the best. The survey results also show that the consumer sample in the survey has similar preferences to the internal consumers, as both groups prefer the internal original image D2 over D1, giving us confidence that the online survey reflects consumer evaluations obtained in the photo booth setting.

\begin{table}[htbp!]
\vspace*{0.2cm}

\caption{Consumer Survey: Choice Probabilities}
\label{table:online_survey}
\centering
\scalebox{0.8}{
\begin{tabular}{lccccc}
\hline 
Choice probabilities   & \makecell{CcGAN with internal \\ and external preferences} & \makecell{CcGAN with \\internal preference} & \makecell{DCGAN} & \makecell{Internal \\ original images} \tabularnewline
\hline 
Model-level   & 33.0\%  & 23.3\%  & 16.0\% & 27.8\% \tabularnewline
\hline 
Image-level   & A1: 24.7\%  & B1: 23.5\%  & C1: 19.2\% & D1: 17.0\% \tabularnewline
              & A2: 45.2\%  & B2: 22.7\%  & C2: 11.4\% & D2: 38.5\% \tabularnewline
              &  A3: 36.0\%  & B3: 23.7\%  & C3: 18.4\% & \tabularnewline
              &  A4: 26.1\%   & B4: 23.1\%  & C4: 15.0\% & \tabularnewline
\hline 
\end{tabular}}
\end{table}

\subsection{Model-predicted popularity}
\label{sec:internal_metric}

As an alternative performance test, we use our trained prediction model to evaluate the performance of the generative models. We focus on the same four sets of images as in the distance-based metric test. For each image, we calculate consumers' choice probabilities using our trained prediction models. To maintain consistency between
the prediction and generation models, we use Prediction Model 1 (Prediction Model 2) to evaluate
the performance of the CcGAN with internal preferences only (with internal and external
preferences). Averaging over the choice probabilities of images within each model, we obtain the model-specific average predicted choice probabilities as shown in Table \ref{table:pred_choice_prob}(a). Choice probabilities of original internal images are normalized to 1 in the first column, with percentage changes shown in other columns. 

The results are consistent with those in  the distance-based metric test. As shown in Table \ref{table:pred_choice_prob}(a), CcGAN with internal and external preferences shows the largest performance improvement over the internal original images (9.13\%) because Prediction Model 2 allows it to fully incorporate popular features from both internal and external sources. CcGAN with internal preference only shows a limited performance improvement (2\%) as Prediction Model 1 is trained solely on internal preferences and cannot fully recognize out-of-the-box features from external templates. DCGAN does not perform well.\footnote{DCGAN does not perform well when evaluated by Prediction Model 1. Although DCGAN incorporates new design features from external templates, Prediction Model 1 cannot recognize the attractiveness of these new features and treats them as unpopular. Thus, DCGAN-generated templates perform worse than the original internal images (-1.61\%). DCGAN performs better (4.47\%) when evaluated by Prediction Model 2, as this model can recognize popular features from external sources in the DCGAN-generated templates.}

Incorporating label weights is important; CcGANs without label weights (Table \ref{table:pred_choice_prob}(b)) show a smaller improvement in predicted choice probabilities compared to CcGANs with label weights (Table \ref{table:pred_choice_prob}(a)).


\begin{table}

\caption{Average Predicted Choice Probabilities}
\label{table:pred_choice_prob}

\begin{centering}
\subfloat[]{
\centering{}%
\scalebox{0.8}{
\begin{tabular}{ccccc}
\hline 
\multirow{2}{*}{Evaluation Methods} & Internal & DCGAN & CcGAN with & CcGAN with Internal\tabularnewline
 & Original Images  &  & Internal Preference & and External Preference\tabularnewline
\hline 
Prediction Model 1 & 1.00 & -1.61\% & 2.00\% & N.A.\tabularnewline
Prediction Model 2 & 1.00 & 4.47\% & N.A. & 9.13\%\tabularnewline
\hline 
\end{tabular}}}
\par\end{centering}
\medskip{}

\begin{centering}
\subfloat[]{
\centering{}%
\scalebox{0.8}{
\begin{tabular}{cccc}
\hline 
\multirow{2}{*}{Evaluation Methods} & Internal & CcGAN with & CcGAN with Internal\tabularnewline
 & Original Images & Internal Preference & and External Preference\tabularnewline
 &  & (no label weights) & (no label weights)\tabularnewline
\hline 
Prediction Model 1 & 1.00 & 0.72\% & N.A.\tabularnewline
Prediction Model 2 & 1.00 & N.A. & 5.37\%\tabularnewline
\hline 
\end{tabular}}}
\par\end{centering}
\medskip{}

{\footnotesize{}Notes: This table presents the average choice probabilities
predicted for real images and images generated using different GAN
models. The first (second) row shows the choice probabilities predicted
by Prediction Model 1 (Prediction Model 2). Choice probabilities of
real internal images are normalized to 1 in the first column, with
percentage changes shown in other columns.}{\footnotesize\par}
\end{table}

\section{Personalized Design Generation}
\label{sec: personalization}

Our proposed framework is well suited for personalized design generation. The CcGAN with both internal and external preferences is trained based on population-wide popularity labels (i.e., the average choice probabilities across consumers). We can use this general CcGAN as a pre-trained model and re-train it using personalized popularity labels to obtain a personalized CcGAN. For a target consumer, we first use the trained predictor to calculate his choice probabilities for each input image, call these probabilities the personalized popularity labels. We then use these personalized labels to guide the re-training of the generative model. The re-trained generative model is now able to conduct personalized design generation by directionally generating new images that is preferred by this specific consumer. 

\begin{figure}[htb!]
\vspace{0.4cm}
\caption{Personalized Design Generation}
\vspace{-0.2cm}
\centering \includegraphics[width=1\textwidth]{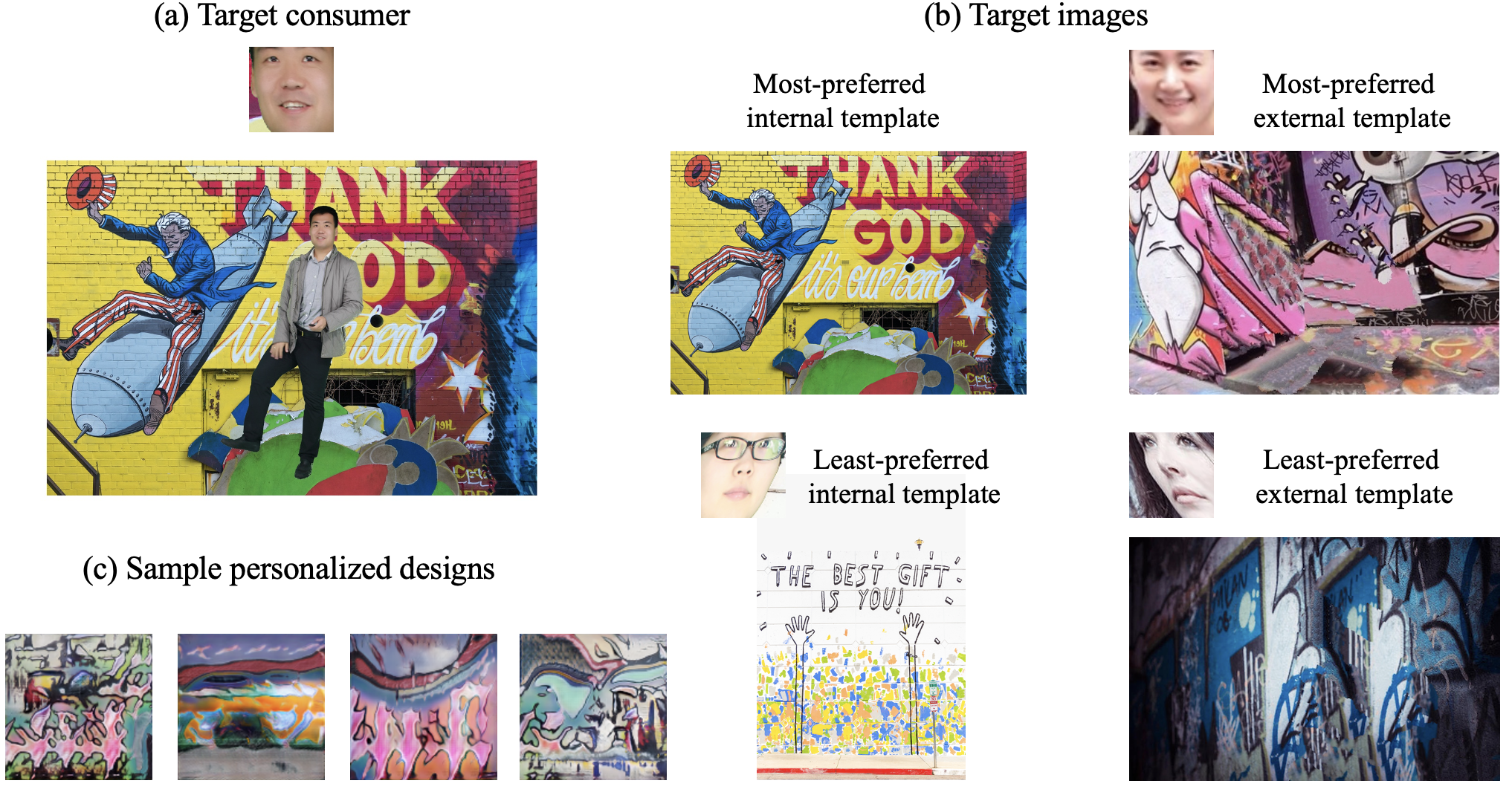}
\label{fig:person} 
\end{figure}

We randomly choose one of the existing internal consumers and conduct a personalization exercise for him. Figure \ref{fig:person}(a) shows the target consumer and the photo he took with the chosen internal template. We use the trained predictor model to calculate his choice probabilities for each input internal and external templates. We then use these probabilities as personalized popularity labels to re-train the general CcGAN for 8000 iterations. This gives us the personalized CcGAN for this consumer. We instruct the personalized CcGAN to generate new templates from the most preferred class. 

To evaluate the performance of the generated personalized templates, we define a personalized distance-based metric by calculating the distance of each generated template to a set of target images that represent the target consumer's most- and least-preferred existing templates. The calculation procedure is similar to the previous distance-based metric test, but the target images are redefined to be personalized: 1) the internal template chosen by the target consumer, representing his most-preferred internal template; 2) the internal template chosen by the internal consumer who looks the least similar to him (based on the distance of their facial appearance embeddings), representing his least-preferred internal template; 3) the external template chosen by the most similar external consumer, representing his most-preferred external template; and 4) the external template chosen by the least similar external consumer, representing his least-preferred external template. Figure \ref{fig:person}(b) displays the four target images. Intuitively, consumers who resemble the focal consumer are likely to share similar preferences, making their choices more appealing to him. Indeed, we find that his most-preferred templates share similar patterns, which are different from his least-preferred templates. The personalized distance-based metric measures how similar the personalized templates are to his most- and least-preferred templates. A smaller metric represents better performance. 

We use the trained personalized CcGAN to generate 100 personalized designs. Figure \ref{fig:person}(c) shows sample personalized designs. Table \ref{table:pred_distance_person} presents the means and medians of the personalized distance-based metrics by the personalized and non-personalized models. We find that the personalized CcGAN with both preferences performs the best, followed by the general CcGAN with both preferences, and the DCGAN performs the worst. The results suggest that incorporating personalized preference yields even better designs than incorporating population-level preference in terms of generating designs that are attractive to a specific consumer.

\begin{table}[htbp!]
\vspace*{0.2cm}

\caption{Personalized Distance-Based Metrics: Mean and Median}
\label{table:pred_distance_person}
\centering
\scalebox{0.8}{
\begin{tabular}{lccc}
\hline 
Statistics  & DCGAN  & \makecell{CcGAN with internal and \\ external preferences} & \makecell{CcGAN with internal and \\ external preferences (personalized)}  \tabularnewline
\hline 
Mean    & 1.22  & 1.10  & 0.97 \tabularnewline
Median  & 1.22  & 1.05  & 0.96 \tabularnewline
\hline 
\end{tabular}}

\raggedright {\footnotesize{}Note: This table presents the mean and median distance-based metrics for images generated by different GAN models. Smaller values indicate the generated templates are visually similar to popular internal templates and distinct from unpopular ones, suggesting higher consumer popularity.}{\footnotesize\par}
\end{table}

\section{Conclusion}
\label{sec:conclusion}

In this paper, we propose a novel framework to transform the product design process in the era of generative AI. The unique contribution of our work is that our proposed model systematically incorporates consumer preferences and external UGC into the design generation process. Two important managerial insights emerge from our results. First, it is beneficial to incorporate consumer preference information into the generation process because it can guide the generator to directionally generate consumer-preferred designs. Second, it is not sufficient to incorporate only external designs; external preference information is also required for the generative model to fully incorporate new appealing features from external designs. 

Our framework offers a superior alternative to the prevailing industry practice that relies solely on pure generative models without considering consumer preferences. The proposed model is preference-aware, cost-effective, automated, and scalable, generating new designs in a controllable and targeted way. The new designs are more likely to be preferred by consumers ex ante, unlike the traditional method of creating many designs and discarding unpopular ones later. Our approach saves resources by avoiding less appealing designs.

Our framework provides an effective solution to the product design ideation problem. Traditional methods depend heavily on professionally trained human designers, who have limited capacity and may struggle to produce high-quality designs on a large scale. Additionally, these designers might not always be fully attuned to the latest consumer demands and often cannot systematically learn the preferences of a broad audience. Our approach enhances the traditional design process by incorporating existing designs from external sources, which can complement human efforts. User-generated content (UGC) images, in particular, can be viewed as real-time design ideas created by individual users, offering a rich and diverse source for ideation.

Our framework provides a cost-effective way to evaluate new product designs by using user-generated content (UGC) instead of traditional consumer surveys. UGC offers a vast, real-time repository of consumer preferences that inform design decisions. This data is highly relevant as it is voluntarily generated by consumers, abundant, readily available online, and constantly updated, helping firms stay aligned with evolving consumer demands. UGC is particularly beneficial for start-ups lacking resources for costly traditional design methods, reducing financial burdens and offering a dynamic, authentic reflection of consumer tastes.

In conclusion, our findings highlight the benefits for industry practitioners of integrating consumer preferences and external UGC into generative models. Unlike pure GAN-based methods, our approach utilizes external preference data from UGC to enhance design processes and extract valuable consumer insights without costly surveys. This framework is widely applicable in artistic, creative, and other domains with diverse consumer preferences. Future research could investigate embedding consumer preferences into generative models across various contexts, expanding understanding of the effectiveness and scalability of these approaches.

\begin{singlespace}
\bibliographystyle{apalike}
\bibliography{reference}
\end{singlespace}
%


\clearpage
\appendix

\end{document}